\documentclass[journal,onecolumn,draftcls,12pt]{IEEEtran}

\usepackage{amssymb, amsmath}
\usepackage{amsfonts}
\usepackage{bbm}
\usepackage{mathrsfs}
\usepackage{xspace}
\usepackage{bm}

\usepackage{cite}
\usepackage{epsfig}
\usepackage{cases}

\newtheorem{theorem}{Theorem}

\newtheorem{proposition}{Prop.}

\newenvironment{textbmatrix}{	\setlength{\arraycolsep}{2.5pt}%
								\big[\begin{matrix}}{\end{matrix}\big]%
								\raisebox{0.08ex}{\vphantom{M}}}

\def\be{\begin{equation}}
\def\ee{\end{equation}}
\def\een{\nonumber \end{equation}}
\def\mat{\begin{bmatrix}}
\def\emat{\end{bmatrix}}
\def\btm{\begin{textbmatrix}}
\def\etm{\end{textbmatrix}}

\def\ba#1\ea{\begin{align}#1\end{align}}
\def\bs#1\es{\begin{split}#1\end{split}} 
\def\bg#1\eg{\begin{gather}#1\end{gather}} 
\def\bi#1\ei{\begin{itemize}#1\end{itemize}}

\newcommand{\safemath}[2]{\newcommand{#1}{\ensuremath{#2}\xspace}}

\DeclareMathOperator{\Tr}{Tr}				
\DeclareMathOperator{\diag}{diag}			
\DeclareMathOperator{\Exop}{\mathbb{E}}		
\DeclareMathOperator{\Varop}{\mathbb{V}\!\mathrm{ar}} 

\safemath{\interior}{\mathrm{Int}}			 
\newcommand{\tp}[1]{\ensuremath{#1^{T}}} 		
\newcommand{\herm}[1]{\ensuremath{#1^{H}}} 	

\safemath{\dfn}{:=}							
\safemath{\dirac}{\delta}					

\safemath{\SNR}{\text{\sc snr}} 				
\safemath{\No}{N_0}							
\safemath{\Es}{E_s}							
\safemath{\Eb}{E_b}							
\safemath{\EbNo}{\frac{\Eb}{\No}}
\safemath{\EsNo}{\frac{\Es}{\No}}

\DeclareMathOperator{\CHop}{\ensuremath{\mathbb{H}}} 
\safemath{\tvir}{h_{\CHop}}					
\safemath{\tvtf}{L_{\CHop}}					
\safemath{\spf}{S_{\CHop}}						
\safemath{\bff}{H_{\CHop}}					

\safemath{\ircf}{R_{h}}						
\safemath{\scf}{R_{S}}						
\safemath{\tfcf}{R_{L}}						
\safemath{\bfcf}{R_{H}}						

\safemath{\mi}{I}							
\safemath{\capacity}{C}						

\safemath{\normal}{\mathcal{N}}				
\safemath{\circnorm}{\mathcal{CN}}			
\safemath{\mchain}{\leftrightarrow}			

\safemath{\dB}{\,\mathrm{dB}}
\safemath{\dBm}{\,\mathrm{dBm}}
\safemath{\Hz}{\,\mathrm{Hz}}
\safemath{\kHz}{\,\mathrm{kHz}}
\safemath{\MHz}{\,\mathrm{MHz}}
\safemath{\GHz}{\,\mathrm{GHz}}
\safemath{\s}{\,\mathrm{s}}
\safemath{\ms}{\,\mathrm{ms}}
\safemath{\mus}{\,\mathrm{\mu s}}
\safemath{\ns}{\,\mathrm{ns}}
\safemath{\meter}{\,\mathrm{m}}
\safemath{\mm}{\,\mathrm{mm}}
\safemath{\cm}{\,\mathrm{cm}}
\safemath{\m}{\,\mathrm{m}}
\safemath{\W}{\,\mathrm{W}}
\safemath{\J}{\,\mathrm{J}}
\safemath{\K}{\,\mathrm{K}}
\safemath{\bit}{\,\mathrm{bit}}

\safemath{\define}{\triangleq}			

\safemath{\equivalent}{\sim}
\safemath{\distas}{\sim}					

\safemath{\reals}{\mathbb{R}}
\safemath{\positivereals}{\mathbb{R}^{+}}
\safemath{\integers}{\mathbb{Z}}
\safemath{\posint}{\mathbb{Z}_{+}}
\safemath{\naturals}{\mathbb{N}}
\safemath{\complexset}{\mathbb{C}}

\safemath{\setA}{\mathcal{A}}
\safemath{\setB}{\mathcal{B}}
\safemath{\setC}{\mathcal{C}}
\safemath{\setD}{\mathcal{D}}
\safemath{\setE}{\mathcal{E}}
\safemath{\setF}{\mathcal{F}}
\safemath{\setG}{\mathcal{G}}
\safemath{\setH}{\mathcal{H}}
\safemath{\setI}{\mathcal{I}}
\safemath{\setJ}{\mathcal{J}}
\safemath{\setK}{\mathcal{K}}
\safemath{\setL}{\mathcal{L}}
\safemath{\setM}{\mathcal{M}}
\safemath{\setN}{\mathcal{N}}
\safemath{\setO}{\mathcal{O}}
\safemath{\setP}{\mathcal{P}}
\safemath{\setQ}{\mathcal{Q}}
\safemath{\setR}{\mathcal{R}}
\safemath{\setS}{\mathcal{S}}
\safemath{\setT}{\mathcal{T}}
\safemath{\setU}{\mathcal{U}}
\safemath{\setV}{\mathcal{V}}
\safemath{\setW}{\mathcal{W}}
\safemath{\setX}{\mathcal{X}}
\safemath{\setY}{\mathcal{Y}}
\safemath{\setZ}{\mathcal{Z}}
\safemath{\emptySet}{\varnothing}

\safemath{\bma}{\mathbf{a}}
\safemath{\bmb}{\mathbf{b}}
\safemath{\bmc}{\mathbf{c}}
\safemath{\bmd}{\mathbf{d}}
\safemath{\bme}{\mathbf{e}}
\safemath{\bmf}{\mathbf{f}}
\safemath{\bmg}{\mathbf{g}}
\safemath{\bmh}{\mathbf{h}}
\safemath{\bmi}{\mathbf{i}}
\safemath{\bmj}{\mathbf{j}}
\safemath{\bmk}{\mathbf{k}}
\safemath{\bml}{\mathbf{l}}
\safemath{\bmm}{\mathbf{m}}
\safemath{\bmn}{\mathbf{n}}
\safemath{\bmo}{\mathbf{o}}
\safemath{\bmp}{\mathbf{p}}
\safemath{\bmq}{\mathbf{q}}
\safemath{\bmr}{\mathbf{r}}
\safemath{\bms}{\mathbf{s}}
\safemath{\bmt}{\mathbf{t}}
\safemath{\bmu}{\mathbf{u}}
\safemath{\bmv}{\mathbf{v}}
\safemath{\bmw}{\mathbf{w}}
\safemath{\bmx}{\mathbf{x}}
\safemath{\bmy}{\mathbf{y}}
\safemath{\bmz}{\mathbf{z}}

\bmdefine{\biad}{a}
\bmdefine{\bibd}{b}
\bmdefine{\bicd}{c}
\bmdefine{\bidd}{d}
\bmdefine{\bied}{e}
\bmdefine{\bifd}{f}
\bmdefine{\bigd}{g}
\bmdefine{\bihd}{h}
\bmdefine{\biid}{i}
\bmdefine{\bijd}{j}
\bmdefine{\bikd}{k}
\bmdefine{\bild}{l}
\bmdefine{\bimd}{m}
\bmdefine{\bind}{n}
\bmdefine{\biod}{o}
\bmdefine{\bipd}{p}
\bmdefine{\biqd}{q}
\bmdefine{\bird}{r}
\bmdefine{\bisd}{s}
\bmdefine{\bitd}{t}
\bmdefine{\biud}{u}
\bmdefine{\bivd}{v}
\bmdefine{\biwd}{w}
\bmdefine{\bixd}{x}
\bmdefine{\biyd}{y}
\bmdefine{\bizd}{z}

\bmdefine{\bixid}{\xi}
\bmdefine{\bilambdad}{\lambda}
\bmdefine{\bimud}{\mu}
\bmdefine{\bithetad}{\theta}
\bmdefine{\biphid}{\phi}

\safemath{\bmia}{\biad}
\safemath{\bmib}{\bibd}
\safemath{\bmic}{\bicd}
\safemath{\bmid}{\bidd}
\safemath{\bmie}{\bied}
\safemath{\bmif}{\bifd}
\safemath{\bmig}{\bigd}
\safemath{\bmih}{\bihd}
\safemath{\bmii}{\biid}
\safemath{\bmij}{\bijd}
\safemath{\bmik}{\bikd}
\safemath{\bmil}{\bild}
\safemath{\bmim}{\bimd}
\safemath{\bmin}{\bind}
\safemath{\bmio}{\biod}
\safemath{\bmip}{\bipd}
\safemath{\bmiq}{\biqd}
\safemath{\bmir}{\bird}
\safemath{\bmis}{\bisd}
\safemath{\bmit}{\bitd}
\safemath{\bmiu}{\biud}
\safemath{\bmiv}{\bivd}
\safemath{\bmiw}{\biwd}
\safemath{\bmix}{\bixd}
\safemath{\bmiy}{\biyd}
\safemath{\bmiz}{\bizd}

\safemath{\bmxi}{\bixid}
\safemath{\bmlambda}{\bilambdad}
\safemath{\bmmu}{\bimud}
\safemath{\bmtheta}{\bithetad}
\safemath{\bmphi}{\biphid}

\safemath{\bA}{\mathbf{A}}
\safemath{\bB}{\mathbf{B}}
\safemath{\bC}{\mathbf{C}}
\safemath{\bD}{\mathbf{D}}
\safemath{\bE}{\mathbf{E}}
\safemath{\bF}{\mathbf{F}}
\safemath{\bG}{\mathbf{G}}
\safemath{\bH}{\mathbf{H}}
\safemath{\bI}{\mathbf{I}}
\safemath{\bJ}{\mathbf{J}}
\safemath{\bK}{\mathbf{K}}
\safemath{\bL}{\mathbf{L}}
\safemath{\bM}{\mathbf{M}}
\safemath{\bN}{\mathbf{N}}
\safemath{\bO}{\mathbf{O}}
\safemath{\bP}{\mathbf{P}}
\safemath{\bQ}{\mathbf{Q}}
\safemath{\bR}{\mathbf{R}}
\safemath{\bS}{\mathbf{S}}
\safemath{\bT}{\mathbf{T}}
\safemath{\bU}{\mathbf{U}}
\safemath{\bV}{\mathbf{V}}
\safemath{\bW}{\mathbf{W}}
\safemath{\bX}{\mathbf{X}}
\safemath{\bY}{\mathbf{Y}}
\safemath{\bZ}{\mathbf{Z}}

\safemath{\bZero}{\mathbf{0}}

\bmdefine{\biAd}{A}
\bmdefine{\biBd}{B}
\bmdefine{\biCd}{C}
\bmdefine{\biDd}{D}
\bmdefine{\biEd}{E}
\bmdefine{\biFd}{F}
\bmdefine{\biGd}{G}
\bmdefine{\biHd}{H}
\bmdefine{\biId}{I}
\bmdefine{\biJd}{J}
\bmdefine{\biKd}{K}
\bmdefine{\biLd}{L}
\bmdefine{\biMd}{M}
\bmdefine{\biOd}{N}
\bmdefine{\biPd}{O}
\bmdefine{\biQd}{P}
\bmdefine{\biRd}{R}
\bmdefine{\biSd}{S}
\bmdefine{\biTd}{T}
\bmdefine{\biUd}{U}
\bmdefine{\biVd}{V}
\bmdefine{\biWd}{W}
\bmdefine{\biXd}{X}
\bmdefine{\biYd}{Y}
\bmdefine{\biZd}{Z}

\bmdefine{\biDelta}{\Delta}
\bmdefine{\biLambda}{\Lambda}
\bmdefine{\biPhi}{\Phi}
\bmdefine{\biSigma}{\Sigma}
\bmdefine{\biOmega}{\Omega}
\bmdefine{\biTheta}{\Theta}

\safemath{\bimA}{\biAd}
\safemath{\bimB}{\biBd}
\safemath{\bimC}{\biCd}
\safemath{\bimD}{\biDd}
\safemath{\bimE}{\biEd}
\safemath{\bimF}{\biFd}
\safemath{\bimG}{\biGd}
\safemath{\bimH}{\biHd}
\safemath{\bimI}{\biId}
\safemath{\bimJ}{\biJd}
\safemath{\bimK}{\biKd}
\safemath{\bimL}{\biLd}
\safemath{\bimM}{\biMd}
\safemath{\bimN}{\biNd}
\safemath{\bimO}{\biOd}
\safemath{\bimP}{\biPd}
\safemath{\bimQ}{\biQd}
\safemath{\bimR}{\biRd}
\safemath{\bimS}{\biSd}
\safemath{\bimT}{\biTd}
\safemath{\bimU}{\biUd}
\safemath{\bimV}{\biVd}
\safemath{\bimW}{\biWd}
\safemath{\bimX}{\biXd}
\safemath{\bimY}{\biYd}
\safemath{\bimZ}{\biZd}

\safemath{\bDelta}{\bielta}
\safemath{\bLambda}{\biLambda}
\safemath{\bPhi}{\biPhi}
\safemath{\bSigma}{\biSigma}
\safemath{\bOmega}{\biOmega}
\safemath{\bTheta}{\biTheta}

\safemath{\veca}{\bma}
\safemath{\vecb}{\bmb}
\safemath{\vecc}{\bmc}
\safemath{\vecd}{\bmd}
\safemath{\vece}{\bme}
\safemath{\vecf}{\bmf}
\safemath{\vecg}{\bmg}
\safemath{\vech}{\bmh}
\safemath{\veci}{\bmi}
\safemath{\vecj}{\bmj}
\safemath{\veck}{\bmk}
\safemath{\vecl}{\bml}
\safemath{\vecm}{\bmm}
\safemath{\vecn}{\bmn}
\safemath{\veco}{\bmo}
\safemath{\vecp}{\bmp}
\safemath{\vecq}{\bmq}
\safemath{\vecr}{\bmr}
\safemath{\vecs}{\bms}
\safemath{\vect}{\bmt}
\safemath{\vecu}{\bmu}
\safemath{\vecv}{\bmv}
\safemath{\vecw}{\bmw}
\safemath{\vecx}{\bmx}
\safemath{\vecy}{\bmy}
\safemath{\vecz}{\bmz}
\safemath{\vecZero}{\bZero}

\safemath{\vecxi}{\bmxi}
\safemath{\veclambda}{\bmlambda}
\safemath{\vecmu}{\bmmu}
\safemath{\vectheta}{\bmtheta}
\safemath{\vecphi}{\bmphi}

\safemath{\matA}{\bA}
\safemath{\matB}{\bB}
\safemath{\matC}{\bC}
\safemath{\matD}{\bD}
\safemath{\matE}{\bE}
\safemath{\matF}{\bF}
\safemath{\matG}{\bG}
\safemath{\matH}{\bH}
\safemath{\matI}{\bI}
\safemath{\matJ}{\bJ}
\safemath{\matK}{\bK}
\safemath{\matL}{\bL}
\safemath{\matM}{\bM}
\safemath{\matN}{\bN}
\safemath{\matO}{\bO}
\safemath{\matP}{\bP}
\safemath{\matQ}{\bQ}
\safemath{\matR}{\bR}
\safemath{\matS}{\bS}
\safemath{\matT}{\bT}
\safemath{\matU}{\bU}
\safemath{\matV}{\bV}
\safemath{\matW}{\bW}
\safemath{\matX}{\bX}
\safemath{\matY}{\bY}
\safemath{\matZ}{\bZ}
\safemath{\matZero}{\bZero}

\safemath{\matDelta}{\bDelta}
\safemath{\matLambda}{\bLambda}
\safemath{\matPhi}{\bPhi}
\safemath{\matSigma}{\bSigma}
\safemath{\matOmega}{\bOmega}
\safemath{\matTheta}{\bTheta}

\safemath{\matIdentity}{\matI}

\newcommand{\utility[2]}{u_{#1}(#2)}
\newcommand{\utilityStar[1]}{u^*_{#1}}

\newcommand{\rate[1]}{R_{#1}}
\safemath{\infobits}{D}
\safemath{\totalbits}{M}
\newcommand{\power[1]}{p_{#1}}
\newcommand{\powerStar[1]}{p^*_{#1}}
\newcommand{\SINR[1]}{\gamma_{#1}}
\newcommand{\SINRStar[1]}{\gamma^*_{#1}}
\newcommand{\efficiencyFunction[1]}{f\left(#1\right)}
\newcommand{\efficiencyFunctionPrime[1]}{f^\prime(#1)}

\newcommand{\channelresponse[2]}{c_{#1}(#2)}
\safemath{\chiptime}{T_c}
\safemath{\srake}{\pathno_S}
\newcommand{\delay[1]}{\tau_{#1}}

\safemath{\SP}{\text{SP}}
\safemath{\SI}{\text{SI}}
\safemath{\MAI}{\text{MAI}}
\newcommand{\channelgain[1]}{h_{#1}}
\newcommand{\hSP[1]}{h_{#1}^{(\SP)}}
\newcommand{\hSI[1]}{h_{#1}^{(\SI)}}
\newcommand{\hMAI[1]}{h_{#1}^{(\MAI)}}
\safemath{\varnoise}{\sigma^2}
\newcommand{\vectornorm}[1]{\left|\left|{#1}\right|\right|}
\newcommand{\vecpathgain[1]}{\bmalpha_{#1}}
\newcommand{\vecrakecoeff[1]}{\bmbeta_{#1}}
\newcommand{\matpathgain[1]}{\matA_{#1}}
\newcommand{\matrakecoeff[1]}{\matB_{#1}}
\newcommand{\matCoeffHsi}{\matPhi}
\newcommand{\coeffHsi[1]}{\phi_{#1}}

\safemath{\game}{G}
\safemath{\userset}{\setK}
\newcommand{\pmin[1]}{\underline{p}_{#1}}
\newcommand{\pmax[1]}{\overline{p}_{#1}}
\newcommand{\powerset[1]}{P_{#1}}
\newcommand{\powervect[1]}{\vecp_{#1}}

\newcommand{\functionGamma[1]}{\Gamma\left({#1}\right)}
\safemath{\powerTimesHsp}{q}
\safemath{\varq}{\sigma^2_\powerTimesHsp}
\safemath{\meanq}{\eta_\powerTimesHsp}

\newcommand{\functionPhi[1]}{\varphi\left({#1}\right)}

\newcommand{\functionTheta[2]}{\theta_{#1}\left({#2}\right)}
\newcommand{\functionThetaQuad[2]}{\theta^2_{#1}\left({#2}\right)}

\safemath{\Po}{P_o}
\newcommand{\nmse[1]}{\text{nmse}\left({#1}\right)}

\newcommand{\functionMu[1]}{\mu\left({#1}\right)}
\newcommand{\functionNu[1]}{\nu\left({#1}\right)}
\newcommand{\functionMuA[1]}{\mu_A\left({#1}\right)}
\newcommand{\functionNuA[1]}{\nu_A\left({#1}\right)}
\newcommand{\varUser[1]}{\sigma^2_{{#1}}}
\newcommand{\stdUser[1]}{\sigma_{{#1}}}
\safemath{\PDPratio}{\Lambda}
\newcommand{\pathgain[2]}{\alpha_{#1}^{(#2)}}
\newcommand{\rakecoeff[2]}{\beta_{#1}^{(#2)}}
\safemath{\pathno}{L}
\safemath{\prake}{\pathno_P}
\safemath{\Pratio}{r}
\safemath{\userno}{K}
\safemath{\frameno}{N_f}
\safemath{\pulseno}{N_c}
\safemath{\gain}{N}
\safemath{\processingMatrix}{\bG}
\newcommand{\MatpathProfile[1]}{\bD^{\alpha}_{#1}}
\newcommand{\MatrakeProfile[1]}{\bD^{\beta}_{#1}}
\newcommand{\MatpathC[1]}{\bC^{\alpha}_{#1}}
\newcommand{\MatrakeC[1]}{\bC^{\beta}_{#1}}
\safemath{\loadFactor}{\rho}
\newcommand{\stepFct[1]}{u\left[{#1}\right]}
\newcommand{\functionVu[1]}{w\left[{#1}\right]}
\newcommand{\SIratio[1]}{\gamma_{0,{#1}}}
\newcommand{\MAIratio[1]}{\zeta_{#1}}
\safemath{\as}{\stackrel{a.s.}{\rightarrow}}
\safemath{\loss}{\Psi}
\newcommand{\distance[1]}{d_{#1}}

\begin{document}

\title{Performance of Rake Receivers in IR-UWB Networks 
Using Energy-Efficient Power Control\thanks{This research 
  was supported in part by the 
  U. S. Air Force Research Laboratory under Cooperative 
  Agreement No. FA8750-06-1-0252, and in part by the 
  U. S. Defense Advanced Research Projects Agency 
  under Grant HR0011-06-1-0052.}}

\author{Giacomo~Bacci\thanks{G. Bacci and M. Luise are with the 
    Dipartimento Ingegneria dell'Informazione, 
    Universit{\`a} di Pisa, Pisa, 56122 Italy
    (e-mail: giacomo.bacci@iet.unipi.it, marco.luise@iet.unipi.it).}, 
  Marco~Luise and H.~Vincent~Poor\thanks{ 
    H. V. Poor is with the Department of Electrical Engineering,
    Princeton University, Princeton, NJ 08544 USA
    (e-mail: poor@princeton.edu).}
}

\maketitle

\begin{abstract}
This paper studies the performance of partial-Rake (PRake) receivers 
in impulse-radio ultrawideband wireless networks when an energy-efficient
power control scheme is adopted. Due to the large bandwidth of the system,
the multipath channel is assumed to be frequency-selective. By making use
of noncooperative game-theoretic models and large-system analysis tools,
explicit expressions are derived in terms of network parameters to measure the 
effects of self-interference and multiple-access interference at a receiving 
access point. Performance of the PRake receivers is thus compared in terms of 
achieved utilities and loss to that of the all-Rake receiver. 
Simulation results are provided to validate the analysis.
\end{abstract}


\section{Introduction}\label{sec:intro}

\PARstart{U}{ltrawideband} (UWB) technology is considered to be a potential
candidate for next-generation multiuser data networks, due to its large
spreading factor (which implies large multiuser capacity) and its lower
spectral density (which allows coexistence with incumbent systems in the same
frequency bands). The requirements for designing high-speed wireless data
terminals include efficient resource allocation and interference reduction.
These issues aim to allow each user to achieve the require quality of 
service (QoS) at the uplink receiver without causing unnecessary interference 
to other users in the system, and minimizing power consumption. 
Energy-efficient power control techniques can be derived making use of 
game theory \cite{mackenzie, goodman, saraydar1, saraydar2, meshkati, bacci}. 
In \cite{mackenzie}, the authors provide motivations for using game theory 
to study power control in communication systems and ad-hoc networks. 
In \cite{goodman}, power control is modeled as a noncooperative game in 
which the users choose their transmit powers to maximize their utilities, 
defined as the ratio of throughput to transmit power. In \cite{saraydar1, 
saraydar2}, the authors use pricing to obtain a more efficient solution for 
the power control game, while the cross-layer problem of joint multiuser 
detection and power control is studied in \cite{meshkati}. A game-theoretic 
approach for a UWB system is studied in \cite{bacci}, where the channel 
fading is assumed to be frequency-selective, due to the large bandwidth 
occupancy \cite{molisch, foerster, molisch2}.

This work extends the results of \cite{bacci}, where a theoretical method to
analyze transmit powers and utilities achieved in the uplink of an 
infrastructure network at the Nash equilibrium has been proposed. However, 
explicit expressions have been derived in \cite{bacci} only for
all-Rake (ARake) receivers \cite{proakis} at the access point, under the 
assumption of a flat averaged power delay profile (aPDP) \cite{hashemi}.
This paper considers partial-Rake (PRake) receivers at the access point and 
makes milder hypotheses on the channel model. Resorting to a large-system 
analysis, we obtain a general characterization of the effects of multiple
access interference (MAI) and self-interference (SI), which allows explicit
expressions for the utilities achieved at the Nash equilibrium to be
derived. Furthermore, we obtain an approximation to the loss of the PRake 
receivers with respect to (wrt) the ARake receivers in terms of 
energy-efficiency, which involves only network parameters and receiver 
characteristics. Since this loss is independent of the channel realizations, 
it can serve as a network design criterion.

The remainder of the paper is organized as follows. Some background for this
work is given in Sect. \ref{sec:background}, where the system model
is described (Sect. \ref{sec:model}) and the results of the game-theoretic
power control approach are shown (Sect. \ref{sec:npcg}). In Sect. 
\ref{sec:interference}, we use a large-system analysis to evaluate the effects
of the interference at the Nash equilibrium. Results are shown for the general 
case, as well as for some particular scenarios (including the one
proposed in \cite{bacci}). Performance of the PRake receivers at the Nash 
equilibrium is analyzed in Sect. \ref{sec:performance}, where also a comparison
with simulation results is provided. Some conclusions are drawn in Sect.
\ref{sec:conclusion}.

\section{Background}\label{sec:background}

\subsection{System Model}\label{sec:model}

Commonly, impulse-radio (IR) systems, which transmit very short pulses with a 
low duty cycle, are employed to implement UWB systems \cite{win}. We focus on a
binary phase shift keying (BPSK) time hopping (TH) IR-UWB system with 
polarity randomization \cite{nakache}. A network with \userno users  
transmitting to a receiver at a common concentration point is
considered. The processing gain of the system is assumed to be 
$\gain=\frameno\cdot\pulseno$, where \frameno is the number of pulses that 
represent one information symbol, and \pulseno denotes the number of possible 
pulse positions in a frame \cite{win}. The transmission is assumed to be 
over \emph{frequency selective channels}, with the channel for user $k$ 
modeled as a tapped delay line:
\be
  \label{eq:channel}
  \channelresponse[k]{t} =
  \sum_{l=1}^{\pathno}{\pathgain[l]{k}\delta(t-(l-1)\chiptime-\delay[k])},
\ee
where \chiptime is the duration of the transmitted UWB pulse, which is
the minimum resolvable path interval; \pathno is the number
of channel paths;
$\vecpathgain[k]=\tp{[\pathgain[1]{k},\dots,\pathgain[\pathno]{k}]}$ and
$\delay[k]$ are the fading coefficients and the delay of user $k$,
respectively. Considering a chip-synchronous scenario, the symbols are
misaligned by an integer multiple of the
chip interval \chiptime: $\delay[k] = \Delta_k\chiptime$, for
every $k$, where $\Delta_k$ is uniformly distributed in
$\{0,1,\dots,\gain-1\}$. In addition we assume that the channel
characteristics remain unchanged over a number of symbol intervals.
This can be justified since the symbol duration in a
typical application is on the order of tens or hundreds of nanoseconds,
and the coherence time of an indoor wireless channel is on the order of
tens of milliseconds.

Due to high resolution of UWB signals, multipath channels can have hundreds 
of multipath components, especially in indoor environments. To mitigate the 
effect of multipath fading as much as possible, we consider an access point 
where \userno Rake receivers\cite{proakis} are used.\footnote{Since the focus
of this work is on the interplay between power control and Rake receivers, 
perfect channel estimation is considered throughout the paper for ease of 
calculation.}
The Rake receiver for user $k$ is in general composed of \pathno coefficients, 
where the vector $\vecrakecoeff[k]=\processingMatrix\cdot\vecpathgain[k]=
\tp{[\rakecoeff[1]{k},\dots,\rakecoeff[\pathno]{k}]}$ represents the combining 
weights for user $k$, and the $\pathno\times\pathno$ matrix 
$\processingMatrix$ depends on the type of Rake receiver employed. 
In particular, if \processingMatrix is a deterministic diagonal matrix, with
\be\label{eq:prakeDef}
  \{\processingMatrix\}_{ll}=
  \begin{cases}
    1, & 1 \le l \le \Pratio\cdot\pathno,\\
    0, & \text{elsewhere},
  \end{cases}
\ee
where $\Pratio\triangleq\prake/\pathno$ and $0<\prake\le\pathno$, a 
PRake with $\prake$ fingers using maximal ratio combining (MRC) is 
considered. It is worth noting that, when $\Pratio=1$, an ARake is implemented.

The signal-to-interference-plus-noise ratio (SINR) of the $k$th user 
at the output of the Rake receiver can be well 
approximated\footnote{This approximation is valid for large \frameno 
(typically, at least 5).} by \cite{gezici1}
\be
  \label{eq:sinr}
  \SINR[k] = \frac{\hSP[k]\power[k]}{\displaystyle{\hSI[k]\power[k] +
      \sum_{\substack{j=1\\j\neq k}}^{\userno}{\hMAI[kj]\power[j]} +
      \sigma^2}},
\ee
where \varnoise is the variance of the additive white Gaussian
noise (AWGN) at the receiver, and the gains are expressed by
\begin{align}
  \label{eq:hSP}
  \hSP[k] &= \herm{\vecrakecoeff[k]}\cdot\vecpathgain[k],\\
  \label{eq:hSI}
  \hSI[k] &= \frac{1}{\gain}
  \frac{\vectornorm{\matCoeffHsi\cdot
      \left(\herm{\matrakecoeff[k]}\cdot\vecpathgain[k]+
      \herm{\matpathgain[k]}\cdot\vecrakecoeff[k]\right)}^2}
       {\herm{\vecrakecoeff[k]}\cdot\vecpathgain[k]},\\
  \intertext{and}
  \label{eq:hMAI}
  \hMAI[kj] &= \frac{1}{\gain}
  \frac{\vectornorm{\herm{\matrakecoeff[k]}\cdot\vecpathgain[j]}^2
  + \vectornorm{\herm{\matpathgain[j]}\cdot\vecrakecoeff[k]}^2
  + \left|\herm{\vecrakecoeff[k]}\cdot\vecpathgain[j]\right|^2}
  {\herm{\vecrakecoeff[k]}\cdot\vecpathgain[k]},
  \intertext{where}
  \label{eq:matrixA}
  \matpathgain[k] &=
  \begin{pmatrix}
    \pathgain[\pathno]{k}&\cdots&\cdots&\pathgain[2]{k}\\
    0&\pathgain[\pathno]{k}&\cdots&\pathgain[3]{k}\\
    \vdots&\ddots&\ddots&\vdots\\
    0&\cdots&0&\pathgain[\pathno]{k}\\
    0&\cdots&\cdots&0
  \end{pmatrix},\\
  \label{eq:matrixB}
  \matrakecoeff[k] &=
  \begin{pmatrix}
    \rakecoeff[\pathno]{k}&\cdots&\cdots&\rakecoeff[2]{k}\\
    0&\rakecoeff[\pathno]{k}&\cdots&\rakecoeff[3]{k}\\
    \vdots&\ddots&\ddots&\vdots\\
    0&\cdots&0&\rakecoeff[\pathno]{k}\\
    0&\cdots&\cdots&0
  \end{pmatrix},\\
  \label{eq:matrixPhi}
  \matCoeffHsi &=
  \diag\left\{\coeffHsi[1],\dots,\coeffHsi[\pathno-1]\right\},\\
  \intertext{and} 
  \label{eq:coeffPhi}
  \coeffHsi[l]&=\sqrt{\frac{\min\{\pathno-l,\pulseno\}}{\pulseno}}
\end{align}
have been introduced for convenience of notation.

\subsection{The Game-Theoretic Power Control Game}\label{sec:npcg}

Consider the application of noncooperative power control techniques to the wireless 
network described above. Focusing on mobile terminals, where it is often more 
important to maximize the number of bits transmitted per Joule of energy 
consumed than to maximize throughput, an energy-efficient approach like the 
one described in \cite{bacci} is considered. 

Game theory \cite{mackenzie} is the natural framework for modeling and 
studying these interactions between users. It is thus possible to consider a 
noncooperative power control game in which each user seeks to maximize
its own utility function as follows. Let $\game = [\userset, \{\powerset[k]\}, 
\{\utility[k]{\powervect[]}\}]$ be the proposed noncooperative game where 
$\userset=\{1,\dots,\userno\}$ is the index set for the users; 
$\powerset[k]=[\pmin[k], \pmax[k]]$ is the strategy set, with 
$\pmin[k]$ and $\pmax[k]$ denoting minimum and maximum power constraints, 
respectively; and $\utility[k]{\powervect[]}$ is the payoff function for 
user $k$ \cite{saraydar2}, defined as
\be
  \label{eq:utility}
  \utility[k]{\powervect[]}=\frac{\infobits}{\totalbits}\rate[k]
  \frac{\efficiencyFunction[{\SINR[k]}]}{\power[k]},
\ee
where $\powervect[]=[\power[1],\dots,\power[\userno]]$ is the vector of 
transmit powers; $\infobits$ and $\totalbits$ are the number of information 
bits per packet and the total number of bits per packet, respectively; 
$\rate[k]$ and $\SINR[k]$ are the transmission rate and the SINR 
(\ref{eq:sinr}) for the $k$th user, 
respectively; and $\efficiencyFunction[{\SINR[k]}]$ is the efficiency function 
representing the packet success rate (PSR), i.e., the probability that a 
packet is received without an error. Throughout this analysis, we assume 
$\pmin[k]=0$ and $\pmax[k]=\pmax[]$ for all $k\in\userset$.

\begin{figure}
  \centering
  \includegraphics[width=12.0cm]{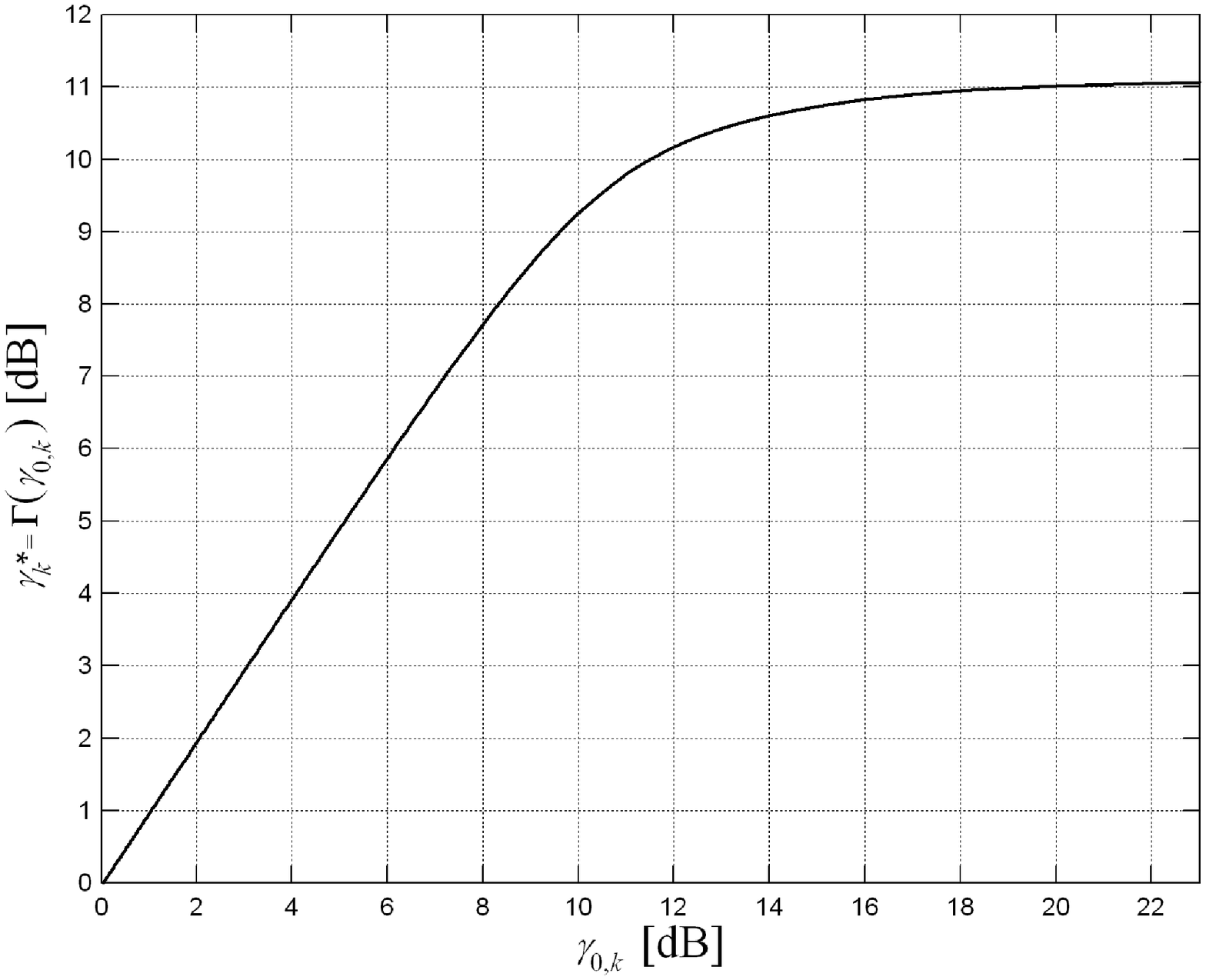}
  \caption{Shape of $\SINRStar[k]$ as a function of $\SIratio[k]$
    ($\totalbits=100$).}
  \label{fig:gammaStar}
\end{figure}

Provided that the efficiency function is increasing, S-shaped, and 
continuously differentiable, with $\efficiencyFunction[0]=0$, 
$\efficiencyFunction[+\infty]=1$, and $\efficiencyFunctionPrime[0]=
d\efficiencyFunction[{\SINR[k]}]/d\SINR[k]|_{\SINR[k]=0}=0$, 
it has been shown \cite{bacci} that the solution of the maximization problem 
$\max_{\power[k]\in\powerset[k]} \utility[k]{\powervect[]}$ for 
$k=1,\dots,\userno$ is
\be
  \label{eq:powerStar}
  \powerStar[k]=\min\left\{
  \frac{\SINRStar[k]\left(\sum_{j\neq k}{\hMAI[kj]\power[j]}+
    \sigma^2\right)}{\hSP[k]\left(1-\SINRStar[k]/\SIratio[k]\right)}, 
  \pmax[]\right\},
\ee
where
\be\label{eq:SIratio}
  \SIratio[k]=\frac{\hSP[k]}{\hSI[k]}
  =\gain\cdot
  \frac{(\herm{\vecrakecoeff[k]}\cdot\vecpathgain[k])^2}
       {\vectornorm{\matCoeffHsi\cdot
           \left(\herm{\matrakecoeff[k]}\cdot\vecpathgain[k]+
           \herm{\matpathgain[k]}\cdot\vecrakecoeff[k]\right)}^2}
       \ge1
\ee
and $\SINRStar[k]$ is the solution of 
\be\label{eq:f_der}
  \efficiencyFunctionPrime[{\SINRStar[k]}]
  \SINRStar[k]\left(1-\SINRStar[k]/\SIratio[k]\right)=
  \efficiencyFunction[{\SINRStar[k]}],
\ee
where $\efficiencyFunctionPrime[{\SINRStar[k]}]=
d\efficiencyFunction[{\SINR[k]}]/d\SINR[k]|_{\SINR[k]=\SINRStar[k]}$.
Since $\SINRStar[k]$ depends only on $\SIratio[k]$, for convenience of 
notation a function $\functionGamma[\cdot]$ is defined such that
$\SINRStar[k]=\functionGamma[{\SIratio[k]}]$. Fig. \ref{fig:gammaStar} 
shows the shape of $\SINRStar[k]=\functionGamma[{\SIratio[k]}]$, where the 
efficiency function is taken to be $\efficiencyFunction[{\SINR[k]}]=
(1-\text{e}^{-\SINR[k]/2})^\totalbits$, with $\totalbits=100$. 

Assuming the typical case of multiuser UWB systems, where $\gain\gg\userno$, 
and also considering $\pmax[]$ sufficiently large, (\ref{eq:powerStar}) can 
be reduced to \cite{bacci}
\be
  \label{eq:minimumPower}
  \powerStar[k]=\frac{1}{\hSP[k]}\cdot
  \frac{\sigma^2\functionGamma[{\SIratio[k]}]}
       {1-\functionGamma[{\SIratio[k]}]\cdot
         \left(\SIratio[k]^{-1}+\MAIratio[k]^{-1}\right)},
\ee
where $\MAIratio[k]^{-1}=\sum_{j\neq k}{\hMAI[kj]/\hSP[j]}$; and 
$\SIratio[k]^{-1}$ is defined as in (\ref{eq:SIratio}).

A necessary and sufficient condition for the Nash equilibrium to be achieved
simultaneously by all \userno users, and thus for (\ref{eq:minimumPower}) 
to be valid, is \cite{bacci}
\be
  \label{eq:requirement}
  \functionGamma[{\SIratio[k]}]\cdot
  \left(\SIratio[k]^{-1}+\MAIratio[k]^{-1}\right)<1
  \quad \forall k\in\userset.
\ee

As can be verified, the amount of transmit power $\powerStar[k]$ required 
to achieve the target SINR $\SINRStar[k]$ will depend not only on the gain 
$\hSP[k]$, but also on the SI term $\hSI[k]$ (through $\SIratio[k]$) and 
the interferers $\hMAI[kj]$ (through $\MAIratio[k]$).

\section{Analysis of the Interference}\label{sec:interference}

In order to derive some quantitative results for the achieved utilities 
and for the transmit powers independent of SI and MAI terms, it is 
possible to resort to a large-system analysis.

\begin{theorem}[\!\!\cite{bacci}]\label{th:oldPaper} 
Assume that $\pathgain[k]{l}$ are zero-mean random variables independent 
across $k$ and $l$, and $\processingMatrix$ is a deterministic diagonal 
matrix (thus implying that $\pathgain[k]{l}$ and $\rakecoeff[j]{m}$ are 
dependent only when $j=k$ and $m=l$). In the asymptotic case where \userno 
and \frameno are finite,\footnote{In order for the analysis to be 
  consistent, and also considering regulations by the US Federal 
  Communications Commission (FCC) \cite{fcc}, it is worth noting that \frameno 
  could not be smaller than a certain threshold ($\frameno\ge5$).} while 
$\pathno, \pulseno \rightarrow \infty$, with the ratio $\pulseno/\pathno$
approaching a constant, the terms $\MAIratio[k]^{-1}$ and 
$\SIratio[k]^{-1}$ converge almost surely (a.s.) to 
\begin{align}
  \label{eq:zetaTh}
  \MAIratio[k]^{-1}&\as
  \frac{1}{\gain}\sum_{\substack{j=1\\j\neq k}}^{\userno}
       {\frac{\functionPhi[{\MatpathProfile[j]\MatrakeC[k]
               \herm{{\MatrakeC[k]}}\MatpathProfile[j]}] +
           \functionPhi[{\MatrakeProfile[k]\MatpathC[j]
               \herm{{\MatpathC[j]}}\MatrakeProfile[k]}]}
         {\functionPhi[{\MatpathProfile[j]\MatrakeProfile[j]}]\cdot
           \functionPhi[{\MatpathProfile[k]\MatrakeProfile[k]}]}}\\
   \intertext{and}
   \label{eq:gammaTh}
   \SIratio[k]^{-1} &\as \frac{1}{\gain}
   \frac{\displaystyle{\lim_{\pathno\rightarrow\infty}{\frac{1}{\pathno^2}
         \sum_{i=1}^{\pathno-1}{\coeffHsi[i]^2}\cdot
         \sum_{l=1}^{i}{\functionThetaQuad[k]{l,\pathno+l-i}}}}}
        {\left(\functionPhi[{\MatpathProfile[k]\MatrakeProfile[k]}]\right)^2},
\end{align}
where $\coeffHsi[i]$ is defined as in (\ref{eq:coeffPhi}); 
$\MatpathProfile[j]$ and $\MatrakeProfile[k]$ are diagonal matrices whose 
elements are 
\begin{align}
  \label{eq:MatpathProfile}
  \{\MatpathProfile[j]\}_{l} &= \sqrt{\Varop[\pathgain[j]{l}]},\\
  \intertext{and}
  \label{eq:MatrakeProfile}
  \{\MatrakeProfile[k]\}_{l} &= \sqrt{\Varop[\rakecoeff[k]{l}]},
\end{align}
with $\Varop[\cdot]$ denoting the variance of a random variable;
$\MatpathC[j]$ and $\MatrakeC[j]$ are $\pathno\times(\pathno-1)$ matrices 
whose elements are 
\begin{align}
  \label{eq:MatpathC}
  \{\MatpathC[j]\}_{li}&=
  \sqrt{\frac{\Varop[\{\matpathgain[j]\}_{li}]}{\pathno}},\\
  \intertext{and}
  \label{eq:MatrakeC}
  \{\MatrakeC[k]\}_{li}&=
  \sqrt{\frac{\Varop[\{\matrakecoeff[k]\}_{li}]}{\pathno}};
\end{align}
$\functionPhi[\cdot]$ is the matrix operator
\be
   \label{eq:Phi}
   \functionPhi[\cdot]=
   \lim_{\pathno\rightarrow\infty}{\frac{1}{\pathno}\Tr(\cdot)}, 
\ee
with $\Tr(\cdot)$ denoting the trace operator; and
\begin{align}\label{eq:Theta}
  \functionTheta[k]{l,\pathno+l-i}=
  \{\MatpathProfile[k]\}_l \{\MatrakeProfile[k]\}_{\pathno+l-i}
  + \{\MatrakeProfile[k]\}_l \{\MatpathProfile[k]\}_{\pathno+l-i}.
\end{align}
\end{theorem}
The proof of this theorem can be found in \cite{bacci}.

The results above can be applied to any kind of fading model, 
as long as the second-order statistics are available. Furthermore, 
due to the symmetry of (\ref{eq:zetaTh}) and (\ref{eq:gammaTh}), 
it is easy to verify that the results are independent of 
large-scale fading models. Hence, Theorem \ref{th:oldPaper}
applies to any kind of channel, which may include both large- 
and small-scale statistics.

\begin{figure}
  \centering
  \includegraphics[width=12.0cm]{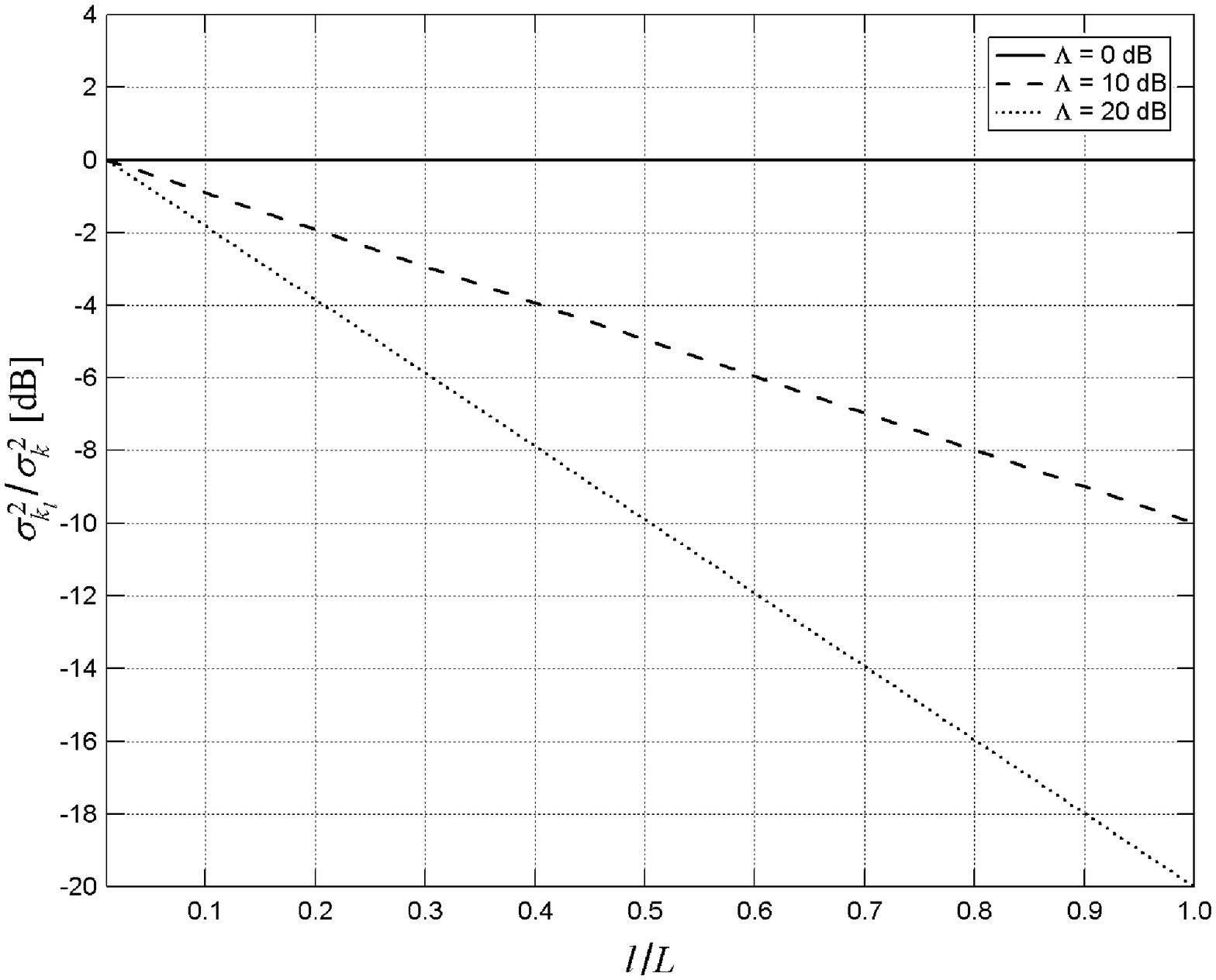}
  \caption{Average power delay profile versus normalized excess delay.}
  \label{fig:pdp}
\end{figure}

Channel modeling for IR-UWB systems is still an open issue. In fact,
while there exists a commonly agreed-on set of basic models for narrowband
and wideband wireless channels \cite{tse}, a similarly well accepted UWB 
channel model does not seem to exist. Recently, two models, namely IEEE 
802.15.3a \cite{foerster} and IEEE 802.15.4a \cite{molisch2}, have been 
standardized to properly characterize the UWB environment. However,
for ease of calculation, the expressions derived in the remainder of the paper 
consider the following simplifying assumptions:

\begin{itemize}
  \item The channel gains are independent complex Gaussian 
    random variables with zero means and variances $\varUser[{k_l}]$, i.e., 
    $\pathgain[k]{l}\distas\circnorm(0, \varUser[{k_l}])$. This assumption 
    leads $|\pathgain[k]{l}|$ to be Rayleigh-distributed with 
    parameter $\varUser[{k_l}]/2$. Although both IEEE 802.15.3a and
    IEEE 802.15.4a models include some forms of Nakagami $m$ distribution 
    for the channel gains, the Rayleigh distribution, appealing for its 
    analytical tractability, has recently been shown \cite{schuster} to 
    provide a good approximation for multipath propagation in UWB systems.

  \item Lately, a clustering phenomenon for the aPDP \cite{hashemi} in 
    IR-UWB multipath channels has emerged from a large number of UWB 
    measurement campaigns \cite{siriwongpairat, chong}. However, owing 
    to the analytical difficulties arising when considering such aspect, 
    this work focuses on an exponentially decaying aPDP, as is customarily 
    used in several UWB channel models \cite{ghassemzadeh, win2}.
    This translates into the hypothesis
    \begin{align}\label{eq:expDecayingPDP}
      \varUser[{k_l}]&=\varUser[k]\cdot\PDPratio^{-\frac{l-1}{\pathno-1}},\\
      \intertext{where}
      \PDPratio&=\varUser[{k_1}]/\varUser[{k_\pathno}]
    \end{align}
    and the variance $\varUser[k]$ depends on the distance between user 
    $k$ and the access point. Fig. \ref{fig:pdp} shows the aPDP for some 
    values of \PDPratio versus the normalized excess delay, i.e., the ratio 
    between the excess delay, $l\chiptime$, and the maximum excess delay 
    considered, $\pathno\chiptime$. It is easy to verify that 
    $\PDPratio=0\,\text{dB}$ represents the case of flat aPDP. 
\end{itemize}

Using these hypotheses, the matrices $\MatpathProfile[k]$ and 
$\MatrakeProfile[k]$ can be expressed in terms of
\begin{align}
  \label{eq:MatpathProfilePRake}
  \{\MatpathProfile[k]\}_l&=\stdUser[k]\cdot
  \PDPratio^{-\frac{l-1}{2(\pathno-1)}}\cdot
  \stepFct[\pathno-l]\\
  \intertext{and}
  \label{eq:MatrakeProfilePRake}
  \{\MatrakeProfile[k]\}_l&=\stdUser[k]\cdot
  \PDPratio^{-\frac{l-1}{2(\pathno-1)}}\cdot
  \stepFct[\Pratio\cdot\pathno-l],
  \intertext{where}
  \label{eq:stepFct}
  \stepFct[n]&=
  \begin{cases}
    1, & n\ge0,\\
    0, & n<0.
  \end{cases}
\end{align}

\subsection{PRake with exponentially decaying aPDP}\label{subsec:prakeExp}

\begin{proposition}\label{th:MAI}
In the asymptotic case where the hypotheses of Theorem \ref{th:oldPaper} hold,
when adopting a PRake with $\prake$ coefficients according to the MRC scheme,
\begin{align}
  \label{eq:thMAI}
  \MAIratio[k]^{-1}&\as\frac{\userno-1}{\gain}\cdot
  \functionMu[{\PDPratio,\Pratio}],
  \intertext{where}
  \label{eq:functionMu}
  \functionMu[{\PDPratio,\Pratio}]&=
  \frac{\left(\PDPratio-1\right)\cdot\PDPratio^{\Pratio-1}}
       {\PDPratio^{\Pratio}-1},
\end{align}
and $\Pratio\triangleq\prake/\pathno$, $0<\Pratio\le1$.
\end{proposition}
The proof can be found in App. \ref{pr:MAI}.

\begin{proposition}\label{th:SI}
In the asymptotic case where the hypotheses of Theorem \ref{th:oldPaper} hold,
when adopting a PRake with $\prake$ coefficients according to the MRC scheme,
\be\label{eq:thSI}
  \SIratio[k]^{-1}\as\frac{1}{\gain}\cdot
  \functionNu[{\PDPratio,\Pratio,\loadFactor}],
\ee
where $\loadFactor\triangleq\pulseno/\pathno$, $0<\loadFactor<\infty$, 
$\Pratio\triangleq\prake/\pathno$, $0<\Pratio\le1$, and 
\begin{subnumcases}{\functionNu[{\PDPratio,\Pratio,\loadFactor}]=}
  \label{eq:nu1st}
  \textstyle{\frac{
      \PDPratio\left(\PDPratio^{\loadFactor}-1\right)
      \left(4\PDPratio^{2\Pratio}+3\PDPratio^{\loadFactor}-1\right)
      -2\PDPratio^{\Pratio+\loadFactor}\left(\PDPratio^{\Pratio}+
      3\PDPratio-1\right)
      \loadFactor\log{\PDPratio}}
    {2\left(\PDPratio^{\Pratio}-1\right)^2\loadFactor
      \PDPratio^{1+\loadFactor}\log{\PDPratio}}},\nonumber\\
  \qquad\qquad
  \text{if $0\le\loadFactor\le\min(\Pratio,1-\Pratio)$};\\
  \label{eq:nu2nd}
  \textstyle{\frac{
      \PDPratio\left(4\PDPratio^{\loadFactor}-1\right)
      \left(\PDPratio^{2\Pratio}-1\right)
      -2\PDPratio^{\Pratio+\loadFactor}
      \left(3\PDPratio\Pratio-\loadFactor+\PDPratio^{\Pratio}
      \loadFactor\right)
      \log{\PDPratio}}
    {2\left(\PDPratio^{\Pratio}-1\right)^2\loadFactor
      \PDPratio^{1+\loadFactor}\log{\PDPratio}}},\nonumber\\
  \qquad\qquad
  \text{if $\min(\Pratio,1-\Pratio)\le\loadFactor\le
    \max(\Pratio,1-\Pratio)$ and 
    $\Pratio\le1/2$};\\
  \label{eq:nu3rd}
  \textstyle{\frac{
      -4\PDPratio^{2+2\Pratio}-4\PDPratio^{2+\loadFactor}+
      \PDPratio^{2(\Pratio+\loadFactor)}+4
      \PDPratio^{2+2\Pratio+\loadFactor}+
      3\PDPratio^{2+2\loadFactor}-2\PDPratio^{1+\Pratio+\loadFactor}
      \left(\Pratio+3\PDPratio\loadFactor+\PDPratio^{\Pratio}
      \loadFactor-1\right)
      \log{\PDPratio}}
    {2\left(\PDPratio^{\Pratio}-1\right)^2\loadFactor
      \PDPratio^{2+\loadFactor}\log{\PDPratio}}},\nonumber\\ 
  \qquad\qquad
  \text{if $\min(\Pratio,1-\Pratio)\le\loadFactor\le
    \max(\Pratio,1-\Pratio)$ and 
    $\Pratio\ge1/2$};\\
  \label{eq:nu4th}
  \textstyle{\frac{
      -\PDPratio^{2+2\Pratio}-4\PDPratio^{2+\loadFactor}+
      \PDPratio^{2(\Pratio+\loadFactor)}+
      4\PDPratio^{2+2\Pratio+\loadFactor}-
      2\PDPratio^{1+\Pratio+\loadFactor}
      \left(\Pratio+3\PDPratio\loadFactor+
      \PDPratio^{\Pratio}\loadFactor-1\right)
      \log{\PDPratio}}
    {2\left(\PDPratio^{\Pratio}-1\right)^2
      \loadFactor\PDPratio^{2+\loadFactor}\log{\PDPratio}}},\nonumber\\ 
  \qquad\qquad
  \text{if $\max(\Pratio,1-\Pratio)\le\loadFactor\le1$};\\
  \label{eq:nu5th}
  \textstyle{\frac{2\PDPratio\left(\PDPratio^{2\Pratio}-1\right)-
      \left(\PDPratio^{\Pratio}+\Pratio+3\PDPratio\Pratio-1\right)
      \PDPratio^{\Pratio}\log{\PDPratio}}
    {\left(\PDPratio^{\Pratio}-1\right)^2
      \loadFactor\PDPratio\log{\PDPratio}}},\nonumber\\ 
  \qquad\qquad
  \text{if $\loadFactor\ge1$}.
\end{subnumcases}
\end{proposition}
The proof can be found in App. \ref{pr:SI}.

Propositions \ref{th:MAI} and \ref{th:SI} give accurate approximations 
for the MAI and SI terms in the general case of PRake receivers at the 
access point and of exponentially decaying aPDP. Furthermore, these
results confirm that the approximations are independent of 
large-scale fading models, as claimed in \cite{bacci}, since they do not 
depend on the variance of the users.

It is also possible to obtain results for more specific scenarios 
using (\ref{eq:thMAI}) and (\ref{eq:thSI}) with particular values 
of $\PDPratio$ and $\Pratio$, as shown in the following subsections.

\subsection{PRake with flat aPDP}\label{subsec:prakeFlat}
The results presented above can be used to study the case of a channel 
model assuming flat aPDP. As already mentioned, the flat aPDP model 
is captured when $\PDPratio=1$. In order to obtain expressions suitable 
for this case, it is sufficient to let $\PDPratio$ go to $1$ in both 
(\ref{eq:thMAI}) and (\ref{eq:thSI}). The former yields
\be
  \lim_{\PDPratio\to1}\functionMu[{\PDPratio,\Pratio}]=
  \frac{1}{\Pratio},
\ee
while the latter gives
\begin{subnumcases}{\lim_{\PDPratio\to1}
    \functionNu[{\PDPratio,\Pratio,\loadFactor}]=}
  \label{eq:nuFlatPrake1st}
  \textstyle{\frac{2\Pratio^2+2\Pratio-4\loadFactor\Pratio+
      \loadFactor^2}{2\Pratio^2}},\nonumber\\
  \qquad\qquad\text{if $0\le\loadFactor\le
    \min(\Pratio,1-\Pratio)$};\\
  \label{eq:nuFlatPrake2nd}
  \textstyle{\frac{1}{2}\left(\frac{2-\loadFactor}{\Pratio}+
    \frac{\Pratio}{\loadFactor}-1\right)},\nonumber\\
  \qquad\qquad\text{if $\min(\Pratio,1-\Pratio)\le
    \loadFactor\le\max(\Pratio,1-\Pratio)$ and 
    $\Pratio\le1/2$};\\
  \label{eq:nuFlatPrake3rd}
  \textstyle{\frac{\Pratio^3+\Pratio^2(9\loadFactor-3)+
      \Pratio(3-9\loadFactor^2)+
      4\loadFactor^3-3\loadFactor^2+
      3\loadFactor-1}{6\loadFactor\Pratio^2}},\nonumber\\
  \qquad\qquad\text{if $\min(\Pratio,1-\Pratio)\le
    \loadFactor\le\max(\Pratio,1-\Pratio)$ and 
    $\Pratio\ge1/2$};\\
  \label{eq:nuFlatPrake4th}
  \textstyle{\frac{4\Pratio^3-3\Pratio^2+3\Pratio+
      (\loadFactor-1)^3}{6\loadFactor\Pratio^2}},\nonumber\\
  \qquad\qquad\text{if $\max(\Pratio,1-\Pratio)\le
    \loadFactor\le1$};\\
  \label{eq:nuFlatPrake5th}
  \textstyle{\frac{4\Pratio^2-3\Pratio+3}{6\loadFactor\Pratio}},\nonumber\\
  \qquad\qquad\text{if $\loadFactor\ge1$}.
\end{subnumcases}

\subsection{ARake with exponentially decaying aPDP}\label{subsec:arakeExp}
The results of Props. \ref{th:MAI}-\ref{th:SI} can also describe
the model of a wireless network using ARake receivers at the 
access point. As noticed in Sect. \ref{sec:model}, an ARake receiver is 
a PRake receiver with $\Pratio=1$. Letting $\Pratio$ go to $1$ in 
(\ref{eq:thMAI}) and (\ref{eq:thSI}), it is possible to obtain 
approximations for the MAI and SI terms in a multipath channel
with exponentially decaying aPDP as follows:
\begin{align}
  \label{eq:muArakeExp}
  \functionMuA[\PDPratio]&=
  \lim_{\Pratio\to1}\functionMu[{\PDPratio,\Pratio}]=1,\\
  \label{eq:nuArakeExp}
  \functionNuA[{\PDPratio,\loadFactor}]&=
  \lim_{\Pratio\to1}\functionNu[{\PDPratio,\Pratio,\loadFactor}]=
  \begin{cases}
    \displaystyle{\frac{2\left(\PDPratio^2-1+\PDPratio^{\loadFactor}-
        \PDPratio^{2-\loadFactor}-
      2\PDPratio\loadFactor\log{\PDPratio}\right)}
         {\left(\PDPratio-1\right)^2\loadFactor\log{\PDPratio}}},
         & \text{if $\loadFactor\le1$},\\
    \displaystyle{\frac{2\left(\PDPratio^2-1-2\PDPratio\log{\PDPratio}\right)}
         {\left(\PDPratio-1\right)^2\loadFactor\log{\PDPratio}}},
         & \text{if $\loadFactor\ge1$}.
  \end{cases}
\end{align}
It is worth noting that the result for $\loadFactor \le 1$ in 
(\ref{eq:nuArakeExp}) has been obtained by letting $\Pratio\to1$ in 
(\ref{eq:nu3rd}).

\subsection{ARake with flat aPDP}\label{subsec:arakeFlat}
The simplest case is represented by a wireless network using the ARake 
receivers at the access point, where the channel is assumed to have a flat 
aPDP. This situation can be captured by simultaneously letting both 
$\PDPratio$ and $\Pratio$ go to $1$ in (\ref{eq:thMAI}) and (\ref{eq:thSI}). 
This approach gives
\begin{align}
  \label{eq:muFlatArake}
  \lim_{\substack{\PDPratio\to1,\\\Pratio\to1}}
  \functionMu[{\PDPratio,\Pratio}]&=1,\\
  \label{eq:nuFlatArake}
  \lim_{\substack{\PDPratio\to1,\\\Pratio\to1}}
  \functionNu[{\PDPratio,\Pratio,\loadFactor}]&=
  \begin{cases}
    \displaystyle{\frac{2}{3}\left(\loadFactor^2-3\loadFactor+3\right)},
    & \text{if $\loadFactor\le1$},\\
    \displaystyle{2/(3\loadFactor)},
    & \text{if $\loadFactor\ge1$}.
  \end{cases}
\end{align}
As in (\ref{eq:nuArakeExp}), the result for $\loadFactor\le1$ in 
(\ref{eq:nuFlatArake}) has been obtained by letting $\Pratio\to1, 
\PDPratio\to1$ in (\ref{eq:nu3rd}).

It is worth noting that (\ref{eq:muFlatArake}) and (\ref{eq:nuFlatArake}) 
coincide with the results obtained in \cite{bacci} for the specific case 
of ARake receivers and flat aPDP.

\subsection{Comments on the Results}

This subsection contains some comments on the results provided by 
Props. \ref{th:MAI}-\ref{th:SI}, applied both to the general case of the
PRake receivers with an exponentially decaying aPDP and to its subcases.

\begin{figure}
  \centering
  \includegraphics[width=12.0cm]{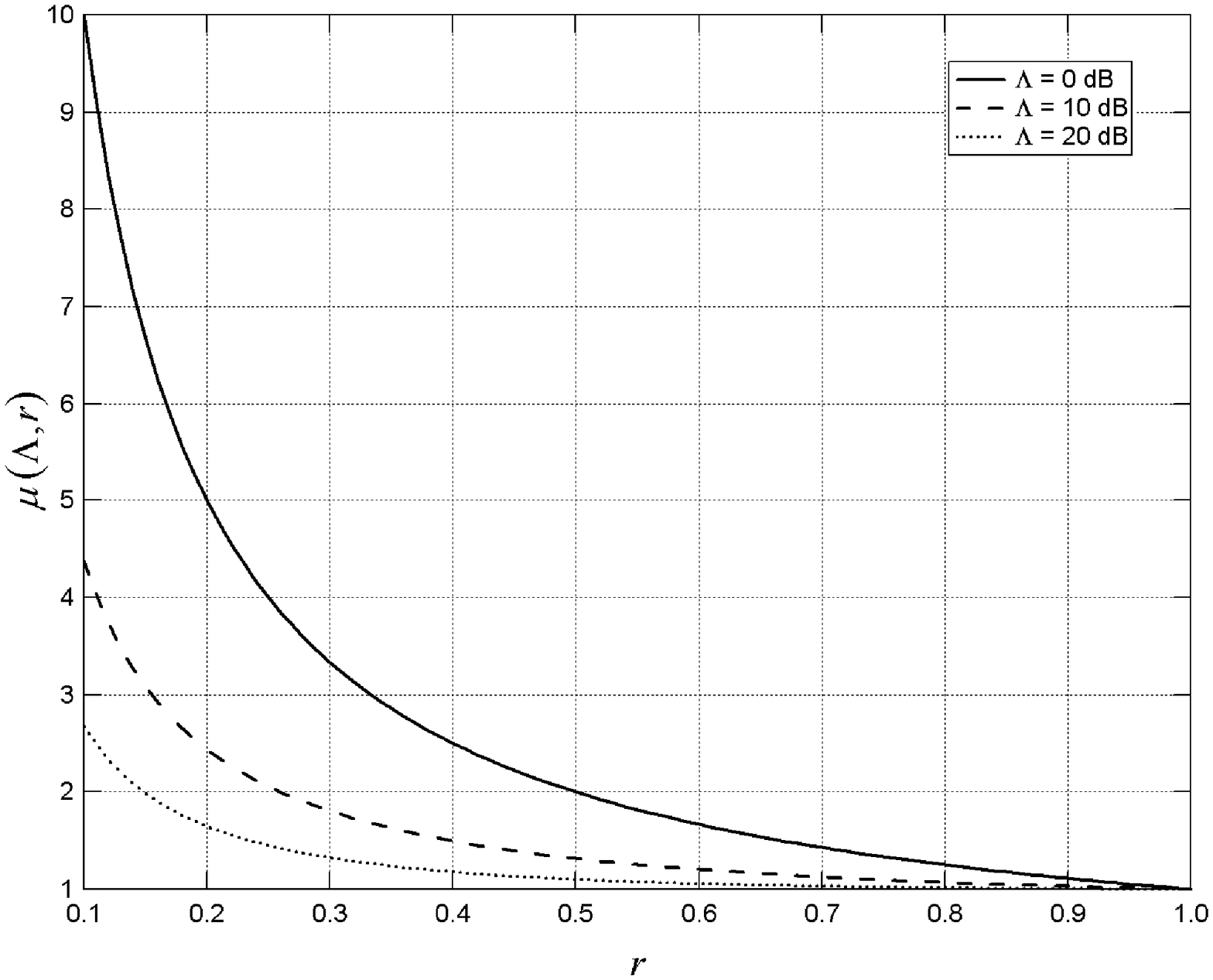}
  \caption{Shape of $\functionMu[{\PDPratio, \Pratio}]$ versus
    $\Pratio$ for some values of $\PDPratio$.}
  \label{fig:mu}
\end{figure}

Fig. \ref{fig:mu} shows the shape of the term 
$\functionMu[{\PDPratio, \Pratio}]$, proportional to the MAI as in 
(\ref{eq:thMAI}), versus the ratio
$\Pratio$ for some values of $\PDPratio$. The solid line represents
$\PDPratio=0\,\text{dB}$, while the dashed and the dotted line depict
$\PDPratio=10\,\text{dB}$ and $\PDPratio=20\,\text{dB}$, respectively.
As can be seen, $\functionMu[{\PDPratio, \Pratio}]$ decreases as 
either $\PDPratio$ or $\Pratio$ increases. Keeping $\Pratio$ fixed, 
it makes sense that $\functionMu[{\PDPratio, \Pratio}]$ is a decreasing 
function of $\PDPratio$, since the received power of the other users is 
lower as $\PDPratio$ increases. Keeping $\PDPratio$ fixed, it makes sense that 
$\functionMu[{\PDPratio, \Pratio}]$ is a decreasing function of $\Pratio$,
since the receiver uses a higher number of coefficients, thus 
better mitigating the effect of MAI. Furthermore, it can be seen that, 
for an ARake, $\lim_{\Pratio\to1}{\functionMu[{\PDPratio, \Pratio}]}=
\functionMuA[\PDPratio]=1$ irrespectively of $\PDPratio$.

\begin{figure}
  \centering
  \includegraphics[width=12.0cm]{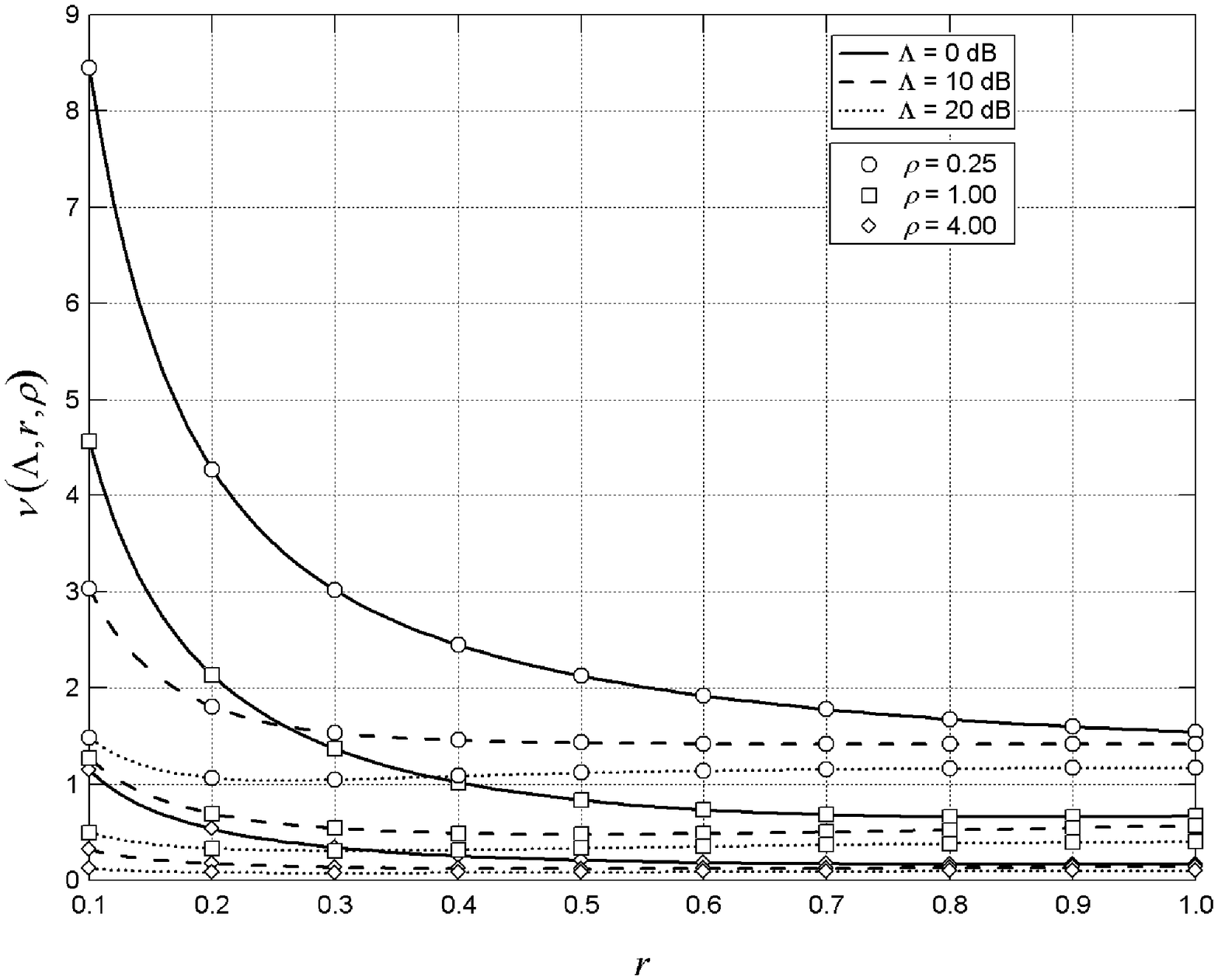}
  \caption{Shape of $\functionNu[{\PDPratio, \Pratio, \loadFactor}]$ 
    versus $\Pratio$ for some values of $\PDPratio$ and $\loadFactor$.}
  \label{fig:nu}
\end{figure}

Fig. \ref{fig:nu} shows the shape of the term
$\functionNu[{\PDPratio, \Pratio, \loadFactor}]$, proportional to the SI as in
(\ref{eq:thSI}), versus the ratio $\Pratio$ for 
some values of $\PDPratio$ and $\loadFactor$. The solid line represents
$\PDPratio=0\,\text{dB}$, while the dashed and the dotted line depict
$\PDPratio=10\,\text{dB}$ and $\PDPratio=20\,\text{dB}$, respectively.
The circles represent $\loadFactor=0.25$, while the square markers and the
rhombi report the shape of $\functionNu[{\PDPratio, \Pratio, \loadFactor}]$ 
for $\loadFactor=1.0$ and $\loadFactor=4.0$, respectively. As can be 
verified, $\functionNu[{\PDPratio, \Pratio, \loadFactor}]$ decreases
as either $\loadFactor$ or $\PDPratio$ increases. This behavior of 
$\functionNu[{\PDPratio, \Pratio, \loadFactor}]$ wrt $\loadFactor$ is
justified by the higher resistance to multipath due to increasing the 
number of possible positions and thus the length of a single frame. This 
also agrees with the results of \cite{bacci} and \cite{gezici1}, where 
it has been shown that, for a fixed total processing gain $\gain$,
systems with higher $\pulseno$ outperform those with smaller $\pulseno$,
due to higher mitigation of SI. Similarly to 
$\functionMu[{\PDPratio, \Pratio}]$, it makes sense that 
$\functionNu[{\PDPratio, \Pratio, \loadFactor}]$ is a 
decreasing function of $\PDPratio$ when $\Pratio$ and $\loadFactor$ are
fixed, since the neglected paths are weaker as $\PDPratio$ increases.
Taking into account the behavior of 
$\functionNu[{\PDPratio, \Pratio, \loadFactor}]$ as a function of $\Pratio$,
it can be verified, either analytically or graphically, that 
$\functionNu[{\PDPratio, \Pratio, \loadFactor}]$ is not 
monotonically decreasing as $\Pratio$ increases. In other words,
an ARake receiver using MRC does not offer the optimum performance in 
mitigating the effect of SI, but it is outperformed by the PRake receivers 
whose $\Pratio$ decreases as $\PDPratio$ increases. This behavior is 
due to the fact that the receiver uses MRC, which attempts 
to gather all the signal energy to 
maximize the signal-to-noise ratio (SNR) and substantially ignores 
the effects of SI \cite{klein}. In this scenario, a minimum mean 
square error (MMSE) combining criterion \cite{gezici2}, while more
complex, might give a different comparison.

\section{Analysis of the Nash Equilibrium}\label{sec:performance}

Making use of the analysis presented in the previous section, it is possible
to study the performance of the PRake receivers in terms of achieved utilities 
when the noncooperative power control techniques described in Sect. 
\ref{sec:npcg} are adopted.

\subsection{Analytical Results}

Using Props. \ref{th:MAI} and \ref{th:SI} in (\ref{eq:utility}) and 
(\ref{eq:minimumPower}), it is straightforward to obtain the expressions 
for transmit powers $\powerStar[k]$ and utilities $\utilityStar[k]$
achieved at the Nash equilibrium, which are independent of the channel 
realizations of the other users, and of SI:
\begin{align}
  \label{eq:power*LSA}
  \powerStar[k]&\as\frac{1}{\hSP[k]}\cdot
  \frac{\gain\sigma^2
    \functionGamma[{\frac{\gain}
      {\functionNu[{\PDPratio,\Pratio,\loadFactor}]}}]}
  {\gain-\functionGamma[{\frac{\gain}
      {\functionNu[{\PDPratio,\Pratio,\loadFactor}]}}]\cdot
    \left[(\userno-1)\functionMu[{\PDPratio,\Pratio}]+
      \functionNu[{\PDPratio,\Pratio,\loadFactor}]\right]},\\
  \label{eq:utilityLSA}
  \utilityStar[k]&\as\hSP[k]\cdot\frac{\infobits}{\totalbits}\rate[k]\cdot
  \efficiencyFunction[{\functionGamma[{\frac{\gain}
      {\functionNu[{\PDPratio,\Pratio,\loadFactor}]}}]}]\cdot
    \frac{\gain-\functionGamma[{\frac{\gain}
        {\functionNu[{\PDPratio,\Pratio,\loadFactor}]}}]\cdot\left[(\userno-1)
        \functionMu[{\PDPratio,\Pratio}]+
        \functionNu[{\PDPratio,\Pratio,\loadFactor}]\right]}
    {\gain\sigma^2\functionGamma[{\frac{\gain}
        {\functionNu[{\PDPratio,\Pratio,\loadFactor}]}}]}.
\end{align}
Note that (\ref{eq:power*LSA})-(\ref{eq:utilityLSA})
require knowledge of the channel realization for user $k$ (through 
$\hSP[k]$).

Analogously, (\ref{eq:requirement}) translates into the system design parameter
\be\label{eq:requirementLSA}
  \frameno\ge\left\lceil\functionGamma[{\frac{\gain}
      {\functionNu[{\PDPratio,\Pratio,\loadFactor}]}}]\cdot
    \frac{(\userno-1)\functionMu[{\PDPratio,\Pratio}]+
      \functionNu[{\PDPratio,\Pratio,\loadFactor}]}
    {\pulseno}\right\rceil,
\ee
where $\lceil\cdot\rceil$ is the ceiling operator.

\begin{proposition}\label{th:loss}
In the asymptotic case where the hypotheses of Theorem \ref{th:oldPaper} hold,
the loss \loss of a PRake receiver wrt an ARake receiver in terms 
of achieved utilities converges a.s. to
\be\label{eq:loss}
  \loss = 
  \frac{\utilityStar[{k_A}]}{\utilityStar[k]}
  \as\functionMu[{\PDPratio,\Pratio}]\cdot
  \frac{\efficiencyFunction[{\functionGamma[{\frac{\gain}
        {\functionNuA[{\PDPratio,\loadFactor}]}}]}]}
  {\efficiencyFunction[{\functionGamma[{\frac{\gain}
        {\functionNu[{\PDPratio,\Pratio,\loadFactor}]}}]}]}
  \cdot
  \frac{\functionGamma[{\frac{\gain}
      {\functionNu[{\PDPratio,\Pratio,\loadFactor}]}}]}
  {\functionGamma[{\frac{\gain}
      {\functionNuA[{\PDPratio,\loadFactor}]}}]}\cdot
  \frac{\gain-\functionGamma[{\frac{\gain}
      {\functionNuA[{\PDPratio,\loadFactor}]}}]
    \left[(\userno-1)\functionMuA[\PDPratio]+
      \functionNuA[{\PDPratio,\loadFactor}]\right]}
  {\gain-\functionGamma[{\frac{\gain}
      {\functionNu[{\PDPratio,\Pratio,\loadFactor}]}}]
    \left[(\userno-1)\functionMu[{\PDPratio,\Pratio}]+
      \functionNu[{\PDPratio,\Pratio,\loadFactor}]\right]},
\ee
where $\utilityStar[{k_A}]$ is the utility achieved by an ARake receiver.
\end{proposition}
The proof can be found in App. \ref{pr:loss}.

Equation (\ref{eq:loss}) also provides a system design criterion. Given 
\pathno, \pulseno, \frameno, \userno and \PDPratio, a desired loss \loss 
can in fact be achieved using the ratio \Pratio obtained by numerically 
inverting (\ref{eq:loss}). Unlike (\ref{eq:power*LSA})-(\ref{eq:utilityLSA}),
this result is independent of all channel realizations.

\subsection{Simulation Results}\label{subsec:simulation}
In this subsection, we show numerical results for the analysis presented in
the previous subsection. Simulations are performed using the iterative 
algorithm described in detail in \cite{bacci}. The systems we examine 
have the design parameters listed in Table \ref{tab:parameters}. 
We use the efficiency function $\efficiencyFunction[{\SINR[k]}]=
(1-\text{e}^{-\SINR[k]/2})^\totalbits$ as a reasonable approximation to 
the PSR \cite{saraydar2, gezici1}. To model the UWB scenario, the 
channel gains are assumed as in Sect. \ref{sec:interference}, with 
$\varUser[k]=0.3\distance[k]^{-2}$, where $\distance[k]$ is the distance 
between the $k$th user and the access point. Distances are assumed to be 
uniformly distributed between $3$ and $20\,\text{m}$. 

\begin{figure}
  \centering
  \includegraphics[width=12.0cm]{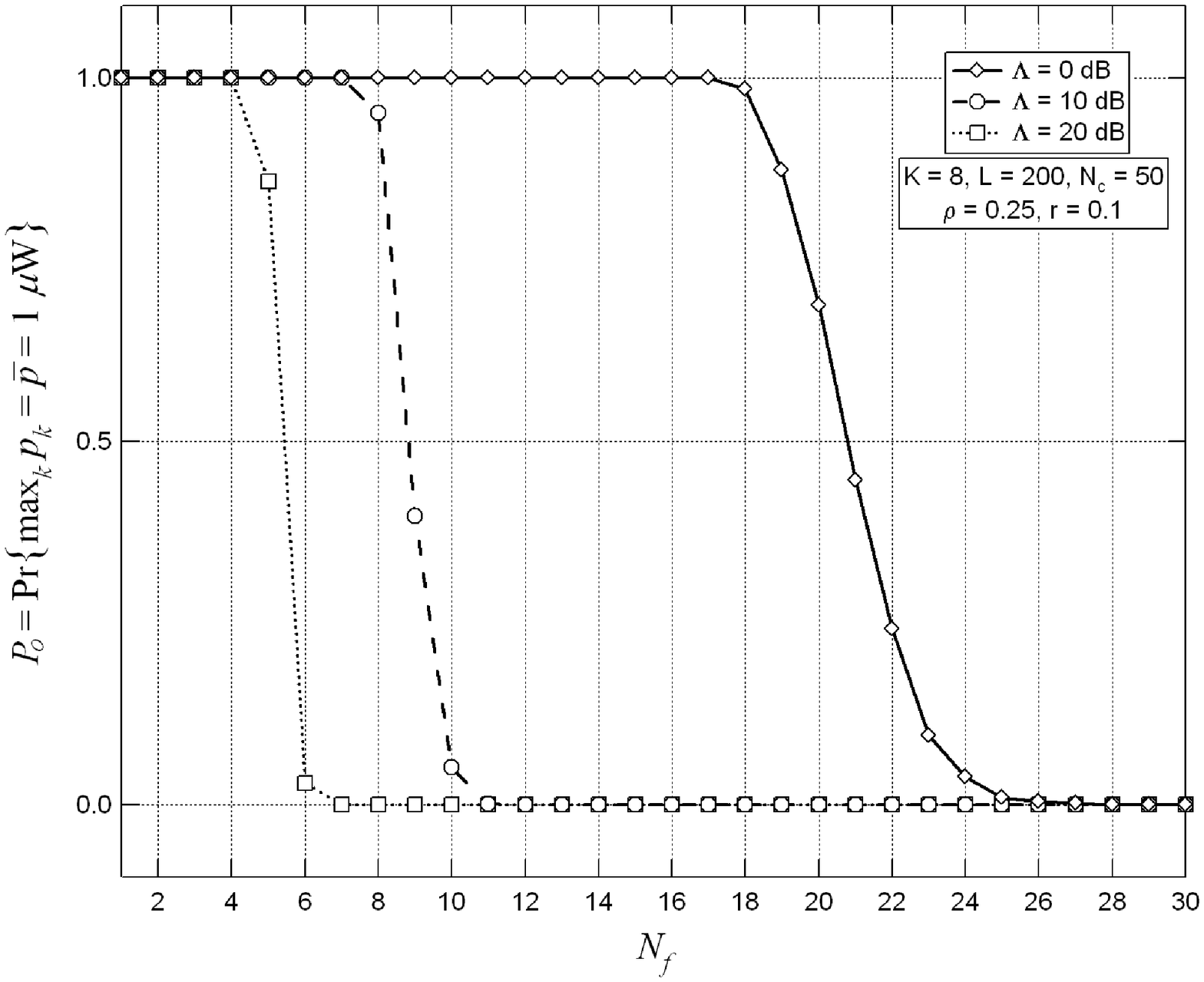}
  \caption{Probability of having at least one user transmitting
    at maximum power versus number of frames.}
  \label{fig:minimumNf}
\end{figure}

Fig. \ref{fig:minimumNf} shows the probability \Po of having at least 
one user transmitting at the maximum power, i.e., 
$\Po=\Pr\{\max_k\power[k]=\pmax[]=1\,\mu\text{W}\}$, as a function of 
the number of frames \frameno. We consider $10\,000$ realizations of the 
channel gains, using a network with $\userno=8$ users, $\pulseno=50$, 
$\pathno=200$ (thus $\loadFactor=0.25$), and PRake receivers with 
$\prake=20$ coefficients (and thus $\Pratio=0.1$). The solid line 
represents the case $\PDPratio=0\,\text{dB}$, while the dashed and the 
dotted lines depict the cases $\PDPratio=10\,\text{dB}$ and
$\PDPratio=20\,\text{dB}$, respectively. Note that the slope of \Po 
increases as $\PDPratio$ increases. This phenomenon is due to reducing 
the effects of neglected path gains as \PDPratio becomes higher, which,
given \frameno, results in having more homogeneous effects of neglected gains. 
Using the parameters above in (\ref{eq:requirementLSA}), the minimum value 
of \frameno that allows all \userno users to simultaneously achieve the 
optimum SINRs is $\frameno=\{21,9,6\}$ for 
$\PDPratio=\{0\,\text{dB}, 10\,\text{dB}, 20\,\text{dB}\}$, respectively.
As can be seen, the analytical results closely match those from simulations. 
It is worth emphasizing that (\ref{eq:requirementLSA}) is valid for both
\pathno and \prake going to $\infty$, as stated in Props. 
\ref{th:MAI}-\ref{th:SI}. In this example, $\prake=20$, which does not 
fulfill this hypothesis. This explains the slight mismatch between theoretical 
and simulation results, especially for small \PDPratio's. However,  
showing numerical results for a feasible system is more interesting than 
simulating a network with a very high number of PRake coefficients.

\begin{figure}
  \centering
  \includegraphics[width=12.0cm]{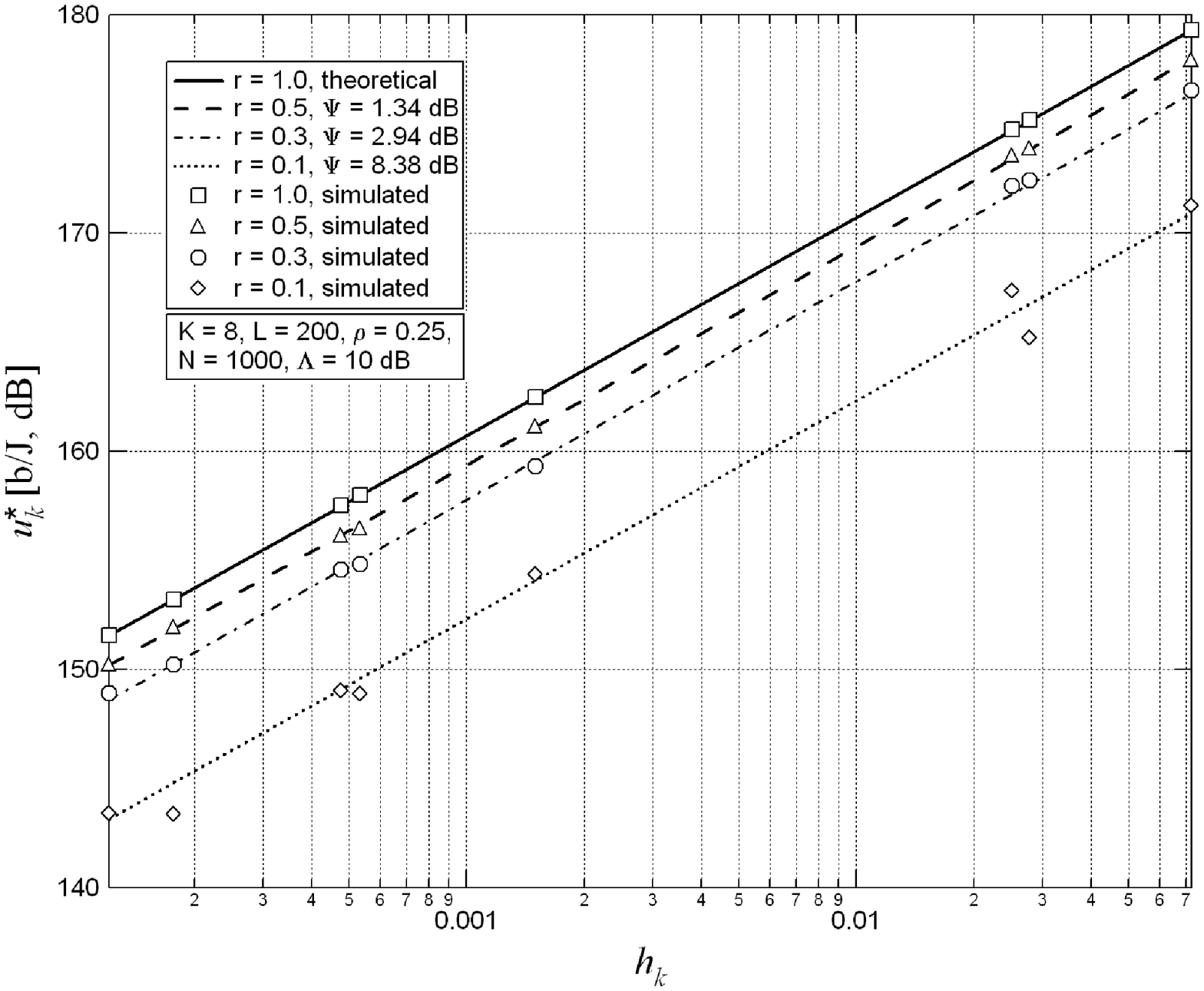}
  \caption{Achieved utility versus channel gain at the Nash equilibrium 
    for different ratios $\Pratio$.}
  \label{fig:loss}
\end{figure}

Fig. \ref{fig:loss} shows a comparison between analytical and numerical
achieved utilities as a function of the channel gains $\channelgain[k]=
\vectornorm{\vecpathgain[k]}^2$. The network has the following parameters: 
$\userno=8$, $\pathno=200$, $\pulseno=50$, $\frameno=20$, 
$\PDPratio=10\,\text{dB}$, $\loadFactor=0.25$. The markers correspond 
to the simulation results given by a single realization of the path gains.
Some values of the number of coefficients of the PRake receiver are 
considered. In particular, the square markers report the results for the 
ARake ($\Pratio=1$), while triangles, circles and rhombi show the cases
$\Pratio=\{0.5,0.3,0.1\}$, respectively. The solid line represents the 
theoretical achieved utility, computed using (\ref{eq:utilityLSA}). The
dashed, the dash-dotted and the dotted lines have been obtained by 
subtracting from (\ref{eq:utilityLSA}) the loss \loss, computed as in 
(\ref{eq:loss}). Using the parameters above, $\loss=\{1.34\,\text{dB}, 
2.94\,\text{dB}, 8.40\,\text{dB}\}$ for $\Pratio=\{0.5,0.3,0.1\}$, 
respectively. As before, the larger the number of $\prake$ coefficients 
is, the smaller the difference between theoretical analysis and simulations is.
It is worth noting that the theoretical results do not consider the actual 
values of $\hSP[k]$, as required in 
(\ref{eq:utilityLSA}),\footnote{This is also valid 
for the case ARake, since $\hSP[k]=\channelgain[k]$.} since they make use of 
the asymptotic approximation (\ref{eq:loss}). As can be verified, the 
analytical results closely match the actual performance of the PRake receivers,
especially recalling that the results are not averaged. Only a single random 
channel realization is in fact considered, because we want to emphasize that 
not only this approximation is accurate on average, but also that the 
normalized mean square error (nmse) 
$\nmse[{\utilityStar[k]}]=\Exop\{\left[\left(\utilityStar[{k_A}]/\loss-
\utilityStar[k]\right)/\utilityStar[k]\right]^2\}$ is considerably low,
where $\Exop\{\cdot\}$ denotes expectation; $\utilityStar[{k_A}]$ and 
\loss are \emph{computed} following (\ref{eq:utilityLSA}) and (\ref{eq:loss}), 
respectively; and $\utilityStar[k]$ represents the \emph{experimental} utility 
at the Nash equilibrium. In fact, by averaging over $10\,000$ channel 
realizations using the same network parameters, 
$\nmse[{\utilityStar[k]}]= \{1.4\times10^{-3}, 5.9\times10^{-3}, 
6.3\times10^{-2}\}$ for $\Pratio=\{0.5, 0.3, 0.1\}$, respectively,
As a conclusion, this allows every network fulfilling the above described 
hypotheses to be studied with the proposed tools. 

\begin{figure}
  \centering
  \includegraphics[width=12.0cm]{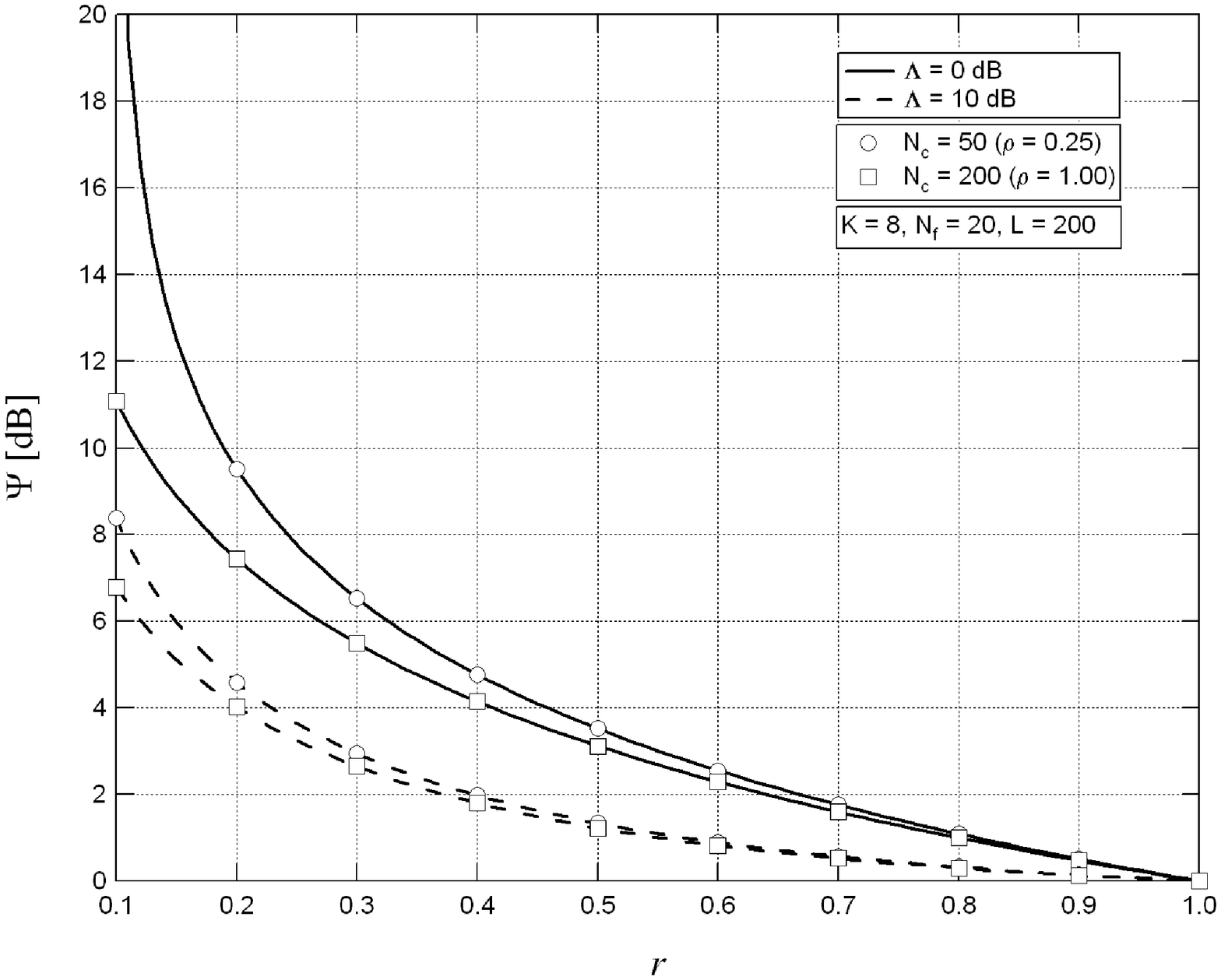}
  \caption{Shape of the loss \loss versus the ratio \Pratio for some values of 
    \PDPratio and \loadFactor.}
  \label{fig:lossVsPratio}
\end{figure}

Fig. \ref{fig:lossVsPratio} shows the loss \loss versus the ratio \Pratio for 
some values of \PDPratio and \loadFactor. The network parameters are set as 
follows: $\userno=8$, $\frameno=20$, and $\pathno=200$. The solid lines 
represent $\PDPratio=0\dB$, while the dashed lines depict $\PDPratio=10\dB$. 
The circles represent $\pulseno=50$ (and thus $\loadFactor=0.25$), while the
square markers report $\pulseno=200$ (and thus $\loadFactor=1.0$). As is 
obvious, \loss is a decreasing function of \Pratio. Furthermore, \loss is a
decreasing function of \PDPratio, since the received power associated to
the paths neglected by the PRake receiver is lower as \PDPratio increases.
Similarly, keeping the number of multiple paths \pathno fixed, \loss decreases
as \loadFactor increases. This complies with theory \cite{bacci, 
gezici1}, since increasing the processing gain provides higher robustness 
against multipath. As a consequence, a system with a lower \loadFactor 
benefits more from a higher number of fingers at the receiver than a system 
with a higher \loadFactor does. Hence, when \loadFactor is lower, a PRake 
receiver performs worse, i.e., \loss is higher. 

It is worth stating that the proposed analysis is mainly 
focused on energy efficiency. Hence, the main performance index here is 
represented by the achieved utility at the Nash equilibrium. However, more
traditional measures of performance such as SINR or bit error rate (BER)
can be obtained using the parameters derived here. In fact, typical
target SINRs at the access point can be computed using $\SINRStar[k]=
\functionGamma[{\gain/\functionNu[{\PDPratio,\Pratio,\loadFactor}]}]$, as
derived in the previous sections. Similarly, the BER can be approximated by 
$Q\left(\sqrt{\SINRStar[k]}\right)$ \cite{gezici1}, where $Q(\cdot)$ denotes 
the complementary cumulative distribution function of a standard normal random 
variable. 

\section{Conclusion}\label{sec:conclusion}

In this paper, we have a used a large-system analysis to study performance
of PRake receivers using maximal ratio combining when energy-efficient
power control techniques are adopted. We have considered a wireless data 
network in frequency-selective environments, where the user terminals
transmit IR-UWB signals to a common concentration point. Assuming the 
averaged power delay profile and the amplitude of the path coefficients to be
exponentially decaying and Rayleigh-distributed, respectively, we have obtained
a general characterization for the terms due to multiple access interference
and self-interference. The expressions are dependent only on the network 
parameters and the number of PRake coefficients. A measure of the loss of 
the PRake receivers with respect to the ARake receiver has then been proposed 
which is completely independent of the channel realizations. This theoretical 
approach may also serve as a criterion for network design, since it is 
completely described by the network parameters. 

\appendix

\subsection{Proof of Prop. \ref{th:MAI}}\label{pr:MAI}

To derive (\ref{eq:thMAI}), we make use of the result (\ref{eq:zetaTh}) of 
Theorem \ref{th:oldPaper}. Using the hypotheses shown in Sect. 
\ref{sec:interference}, $\MatpathProfile[k]$ and $\MatrakeProfile[k]$ are 
represented by (\ref{eq:MatpathProfilePRake}) and 
(\ref{eq:MatrakeProfilePRake}), respectively.

Hence, focusing on the denominator of (\ref{eq:zetaTh}),
\begin{align}
  \label{eq:denMAI1}
  \functionPhi[{\MatpathProfile[k]\MatrakeProfile[k]}] &=
  \lim_{\pathno\to\infty}\frac{1}{\pathno}
  \sum_{l=1}^{\pathno}{\{\MatpathProfile[k]\MatrakeProfile[k]\}_l}
  =\lim_{\pathno\to\infty}\frac{\varUser[k]}{\pathno}
  \sum_{l=1}^{\Pratio\pathno}{\PDPratio^{-\frac{l-1}{\pathno-1}}}
  =\varUser[k]\cdot
  \frac{\PDPratio^{\Pratio}-1}{\PDPratio^{\Pratio}\log{\PDPratio}}.\\
  \intertext{Analogously,}
  \label{eq:denMAI2}
  \functionPhi[{\MatpathProfile[j]\MatrakeProfile[j]}] &=
  \varUser[j]\cdot\frac{\PDPratio^{\Pratio}-1}{\PDPratio^{\Pratio}
    \log{\PDPratio}}.
\end{align}

Using (\ref{eq:matrixA}), (\ref{eq:matrixB}) and (\ref{eq:expDecayingPDP}), 
after some algebraic manipulation, we obtain
\begin{align}
  \{\MatpathC[j]\herm{{\MatpathC[j]}}\}_{ll}&=
  \frac{\varUser[j]}{\pathno}\left(\sum_{m=l+1}^{\pathno}
       {\PDPratio^{-\frac{m-1}{\pathno-1}}}\right)
       \stepFct[\pathno-1-l],\\
  \{\MatrakeC[k]\herm{{\MatrakeC[k]}}\}_{ll}&=
  \frac{\varUser[k]}{\pathno}\left(\sum_{m=l+1}^{\Pratio\pathno}
       {\PDPratio^{-\frac{m-1}{\pathno-1}}}\right)
       \stepFct[\Pratio\pathno-1-l],
\end{align}
where $\stepFct[\cdot]$ is defined as in (\ref{eq:stepFct}). The terms in the 
numerator of (\ref{eq:zetaTh}) thus translate into
\begin{align}
  \label{eq:numMAI1}
  \functionPhi[{\MatpathProfile[j]\MatrakeC[k]
      \herm{{\MatrakeC[k]}}\MatpathProfile[j]}] &=
  \lim_{\pathno\to\infty}\frac{1}{\pathno}
  \sum_{l=1}^{\pathno}{\{\MatpathProfile[j]\}^2_l\{\MatrakeC[k]
    \herm{{\MatrakeC[k]}}\}_{ll}}\nonumber\\
  &=\lim_{\pathno\to\infty}\frac{\varUser[k]\varUser[j]}{\pathno^2}
  \sum_{l=1}^{\Pratio\pathno-1}{\PDPratio^{-\frac{l-1}{\pathno-1}}
    \sum_{m=l+1}^{\Pratio\pathno}
        {\PDPratio^{-\frac{m-1}{\pathno-1}}}}
  =\varUser[k]\varUser[j]\cdot\frac{\PDPratio^{-2\Pratio}
    \left(\PDPratio^{\Pratio}-1\right)^2}
      {2\left(\log{\PDPratio}\right)^2}\\
  \intertext{and}
  \label{eq:numMAI2}
  \functionPhi[{\MatrakeProfile[k]\MatpathC[j]
      \herm{{\MatpathC[j]}}\MatrakeProfile[k]}] &=
  \lim_{\pathno\to\infty}\frac{1}{\pathno}
  \sum_{l=1}^{\pathno}{\{\MatrakeProfile[k]\}^2_l\{\MatpathC[j]
    \herm{{\MatpathC[j]}}\}_{ll}}
  =\lim_{\pathno\to\infty}\frac{\varUser[k]\varUser[j]}{\pathno^2}
  \sum_{l=1}^{\Pratio\pathno}{\PDPratio^{-\frac{l-1}{\pathno-1}}
    \sum_{m=l+1}^{\pathno}{\PDPratio^{-\frac{m-1}{\pathno-1}}}}\nonumber\\
  &=\varUser[k]\varUser[j]\cdot\frac{\PDPratio^{-1-2\Pratio}
    \left(\PDPratio^{\Pratio}-1\right)
    \left(\PDPratio-2\PDPratio^{\Pratio}+\PDPratio^{\Pratio+1}\right)}
      {2\left(\log{\PDPratio}\right)^2}.
\end{align}
Using (\ref{eq:denMAI1})-(\ref{eq:denMAI2}) and 
(\ref{eq:numMAI1})-(\ref{eq:numMAI2}),
\begin{align}
  \label{eq:thMAIend}
  \frac{\hMAI[kj]}{\hSP[j]}\as\frac{1}{\gain}\cdot
  \frac{\functionPhi[{\MatpathProfile[j]\MatrakeC[k]
        \herm{{\MatrakeC[k]}}\MatpathProfile[j]}] +
    \functionPhi[{\MatrakeProfile[k]\MatpathC[j]
        \herm{{\MatpathC[j]}}\MatrakeProfile[k]}]}
       {\functionPhi[{\MatpathProfile[j]\MatrakeProfile[j]}]\cdot
         \functionPhi[{\MatpathProfile[k]\MatrakeProfile[k]}]}
       =\frac{1}{\gain}\cdot
       \frac{\left(\PDPratio-1\right)\PDPratio^{\Pratio-1}}
            {\PDPratio^{\Pratio}-1}.
\end{align}
Using (\ref{eq:thMAIend}), the result (\ref{eq:thMAI}) is straightforward.

\subsection{Proof of Prop. \ref{th:SI}}\label{pr:SI}

To derive (\ref{eq:thSI}), we make use of the result (\ref{eq:gammaTh}) of 
Theorem \ref{th:oldPaper}. Using the hypotheses shown in Sect. 
\ref{sec:interference}, $\MatpathProfile[k]$ and $\MatrakeProfile[k]$ are 
represented by (\ref{eq:MatpathProfilePRake}) and 
(\ref{eq:MatrakeProfilePRake}), respectively. The denominator can be obtained 
following the same steps as in App. \ref{pr:MAI}:
\be\label{eq:denSI}
  \left(\functionPhi[{\MatpathProfile[k]\MatrakeProfile[k]}]\right)^2=
  \stdUser[k]^4\cdot\frac{\left(\PDPratio^{\Pratio}-1\right)^2}
          {\PDPratio^{2\Pratio}\left(\log{\PDPratio}\right)^2}.
\ee
Following (\ref{eq:Theta}), 
\be
  \functionThetaQuad[k]{{l,\pathno+l-i}}=
  \stdUser[k]^4\cdot\PDPratio^{-\frac{\pathno+2l-i-2}{\pathno-1}}
  \cdot\functionVu[{l,i}],
\ee
where
\begin{align}
  \functionVu[{l,i}]&=\stepFct[{\Pratio\pathno-l}]+
  \stepFct[{\Pratio\pathno-\pathno+i-l}]\nonumber\\
  &+2\stepFct[{\Pratio\pathno-l}]\cdot\stepFct[{\Pratio\pathno-\pathno+i-l}]
\end{align}
has been introduced for convenience of notation.

In order to obtain explicit expressions for $\functionVu[{l,i}]$, it is 
convenient to split the range of \Pratio into the two following cases.
\begin{itemize}
  \item $\Pratio\le1/2$: taking into account all the possible values of 
    $l$ and $i$, 
    \be\label{eq:vu1st}
      \functionVu[{l,i}]=
      \begin{cases}
        4, & \text{if $\pathno-\Pratio\pathno+1 \le i \le \pathno-1$ and
          $1 \le l \le \Pratio\pathno-\pathno+i$};\\
        1, & \text{either if $1 \le i \le \Pratio\pathno$ and 
          $1 \le l \le 1$,}\\
        & \text{or if $\Pratio\pathno \le i \le \pathno-\Pratio\pathno$ 
          and $1 \le l \le\Pratio\pathno$},\\
        & \text{or if $\pathno-\Pratio\pathno+1 \le i \le \pathno-1$ and
          $\Pratio\pathno-\pathno+i+1 \le 
          l \le \Pratio\pathno$};\\
        0, & \text{elsewhere}.
      \end{cases}
    \ee
    Substituting (\ref{eq:Theta}) and (\ref{eq:vu1st}) in the numerator of 
    (\ref{eq:gammaTh}) yields
    \begin{align}
      \label{eq:numSI1}
      &\frac{1}{\stdUser[k]^4}
      \sum_{i=1}^{\pathno-1}{\coeffHsi[i]^2}\cdot
      \sum_{l=1}^{i}{\functionThetaQuad[k]{l,\pathno+l-i}}=\nonumber\\
      &=\sum_{i=1}^{\Pratio\pathno}{\coeffHsi[i]^2}\cdot
      \sum_{l=1}^{i}{\PDPratio^{-\frac{\pathno+2l-i-2}{\pathno-1}}}
      +\sum_{i=\Pratio\pathno+1}^{\pathno-\Pratio\pathno}
          {\coeffHsi[i]^2}\cdot
      \sum_{l=1}^{\Pratio\pathno}{\PDPratio^{-\frac{\pathno+2l-i-2}
          {\pathno-1}}}\nonumber\\
      &+\sum_{i=\pathno-\Pratio\pathno+1}^{\pathno-1}
          {\coeffHsi[i]^2}\cdot
      \sum_{l=1}^{\Pratio\pathno-\pathno+i}
          {4\PDPratio^{-\frac{\pathno+2l-i-2}{\pathno-1}}}
      +\sum_{i=\pathno-\Pratio\pathno+1}^{\pathno-1}
          {\coeffHsi[i]^2}\cdot
      \sum_{l=\Pratio\pathno-\pathno+i+1}^{\Pratio\pathno}
          {\PDPratio^{-\frac{\pathno+2l-i-2}{\pathno-1}}};
    \end{align}

    \item $\Pratio\ge1/2$: taking into account all the possible values 
      of $l$ and $i$, 
    \be\label{eq:vu2nd}
      \functionVu[{l,i}]=
      \begin{cases}
        4, & \text{either if $\pathno-\Pratio\pathno+1 \le 
          i \le \Pratio\pathno$ and
          $1 \le l \le \Pratio\pathno-\pathno+i$},\\
        & \text{or if $\Pratio\pathno+1 \le i \le \pathno-1$ and
          $1 \le l \le \Pratio\pathno-\pathno+i$};\\
        1, & \text{either if $1 \le i \le \pathno-\Pratio\pathno$ and 
          $1 \le l \le 1$,}\\
           & \text{or if $\pathno-\Pratio\pathno+1 \le i \le 
          \Pratio\pathno$ and
          $\Pratio\pathno-\pathno+i+1 \le l \le i$},\\
           & \text{or if $\Pratio\pathno+1 \le i \le \pathno-1$ and
             $\Pratio\pathno-\pathno+i+1 \le l 
             \le \Pratio\pathno$};\\
        0, & \text{elsewhere}.
      \end{cases}
    \ee
    Substituting (\ref{eq:Theta}) and (\ref{eq:vu2nd}) in the numerator of 
    (\ref{eq:gammaTh}) yields
    \begin{align}
      \label{eq:numSI2}
      &\frac{1}{\stdUser[k]^4}
      \sum_{i=1}^{\pathno-1}{\coeffHsi[i]^2}\cdot
      \sum_{l=1}^{i}{\functionThetaQuad[k]{l,\pathno+l-i}}=
      \sum_{i=1}^{\pathno-\Pratio\pathno}{\coeffHsi[i]^2}\cdot
      \sum_{l=1}^{i}{\PDPratio^{-\frac{\pathno+2l-i-2}{\pathno-1}}}\nonumber\\
      &+\sum_{i=\pathno-\Pratio\pathno+1}^{\Pratio\pathno}
          {\coeffHsi[i]^2}\cdot
      \sum_{l=1}^{\Pratio\pathno-\pathno+i}
          {4\PDPratio^{-\frac{\pathno+2l-i-2}{\pathno-1}}}
      +\sum_{i=\pathno-\Pratio\pathno+1}^{\Pratio\pathno}
          {\coeffHsi[i]^2}\cdot
      \sum_{l=\Pratio\pathno-\pathno+i+1}^{i}
          {\PDPratio^{-\frac{\pathno+2l-i-2}{\pathno-1}}}\nonumber\\
      &+\sum_{i=\Pratio\pathno+1}^{\pathno-1}{\coeffHsi[i]^2}\cdot
      \sum_{l=1}^{\Pratio\pathno-\pathno+i}
          {4\PDPratio^{-\frac{\pathno+2l-i-2}{\pathno-1}}}
      +\sum_{i=\Pratio\pathno+1}^{\pathno-1}{\coeffHsi[i]^2}\cdot
      \sum_{l=\Pratio\pathno-\pathno+i+1}^{\Pratio\pathno}
          {\PDPratio^{-\frac{\pathno+2l-i-2}{\pathno-1}}}.
    \end{align}

\end{itemize}

In order to obtain (\ref{eq:nu1st})-(\ref{eq:nu5th}), the explicit values of 
$\coeffHsi[i]^2$ must be used. From (\ref{eq:matrixPhi})-(\ref{eq:coeffPhi}) 
follows 
\be\label{eq:coefHpr}
  \coeffHsi[i]^2=
  \begin{cases}
    (\pathno-i)/\pulseno, & \text{either if $\pulseno\le\pathno$ and
      $\pathno-\pulseno+1 \le i \le \pathno-1$},\\
    & \text{or if $\pulseno\ge\pathno$ and $1 \le i \le \pathno-1$};\\
    1, & \text{if $\pulseno\le\pathno$ and $1 \le i \le \pathno-\pulseno$}.
  \end{cases}
\ee
As in the case of \Pratio, it is convenient to separate the range of 
$\loadFactor=\pulseno/\pathno$ in the following cases.
\begin{itemize}
  \item $0 \le \loadFactor \le \min(\Pratio, 1-\Pratio)$: substituting 
    (\ref{eq:coefHpr}) in (\ref{eq:numSI1}) and (\ref{eq:numSI2}), they 
    both yield
    \begin{align}\label{eq:prNu1st}
      \frac{1}{\stdUser[k]^4}
      \lim_{\pathno\rightarrow\infty}{\frac{1}{\pathno^2}
        \sum_{i=1}^{\pathno-1}{\coeffHsi[i]^2}\cdot
        \sum_{l=1}^{i}{\functionThetaQuad[k]{l,\pathno+l-i}}}&=
      \frac{\PDPratio\left(\PDPratio^{\Pratio}-1\right)
      \left(4\PDPratio^{2\Pratio}+3\PDPratio^{\loadFactor}-1\right)}
      {2\PDPratio^{\loadFactor+2\Pratio+1}\loadFactor
        \left(\log{\PDPratio}\right)^3}\nonumber\\
      &-\frac{2\PDPratio^{\Pratio+\loadFactor}
        \left(\PDPratio^{\Pratio}+3\PDPratio-1\right)
      \loadFactor\log{\PDPratio}}
           {2\PDPratio^{\loadFactor+2\Pratio+1}\loadFactor
             \left(\log{\PDPratio}\right)^3}.
    \end{align}
    Making use of (\ref{eq:gammaTh}), (\ref{eq:denSI}) and (\ref{eq:prNu1st}),
    the results (\ref{eq:thSI}) and (\ref{eq:nu1st}) are straightforward.
    
  \item $\min(\Pratio, 1-\Pratio) \le \loadFactor\le 
    \max(\Pratio, 1-\Pratio)$ and $\Pratio\le 1/2$: substituting 
    (\ref{eq:coefHpr}) in (\ref{eq:numSI1}) yields
    \begin{align}\label{eq:prNu2nd}
      \frac{1}{\stdUser[k]^4}
      \lim_{\pathno\rightarrow\infty}{\frac{1}{\pathno^2}
        \sum_{i=1}^{\pathno-1}{\coeffHsi[i]^2}\cdot
        \sum_{l=1}^{i}{\functionThetaQuad[k]{l,\pathno+l-i}}}&=
      \frac{\PDPratio\left(\PDPratio^{2\Pratio}-1\right)
      \left(4\PDPratio^{\loadFactor}-1\right)}
      {2\PDPratio^{\loadFactor+2\Pratio+1}\loadFactor
        \left(\log{\PDPratio}\right)^3}\nonumber\\
      &-\frac{2\PDPratio^{\Pratio+\loadFactor}
        \left(3\PDPratio\Pratio-\loadFactor+\PDPratio^{\Pratio}
        \loadFactor\right)\log{\PDPratio}}
      {2\PDPratio^{\loadFactor+2\Pratio+1}\loadFactor
        \left(\log{\PDPratio}\right)^3}.
    \end{align}
    Making use of (\ref{eq:gammaTh}), (\ref{eq:denSI}) and (\ref{eq:prNu2nd}),
    the results (\ref{eq:thSI}) and (\ref{eq:nu2nd}) are straightforward.
    
  \item $\min(\Pratio, 1-\Pratio) \le \loadFactor\le 
    \max(\Pratio, 1-\Pratio)$ and $\Pratio\ge 1/2$: substituting 
    (\ref{eq:coefHpr}) in (\ref{eq:numSI2}) yields
    \begin{align}\label{eq:prNu3rd}
      \frac{1}{\stdUser[k]^4}\lim_{\pathno\rightarrow\infty}
      {\frac{1}{\pathno^2}
        \sum_{i=1}^{\pathno-1}{\coeffHsi[i]^2}\cdot
        \sum_{l=1}^{i}{\functionThetaQuad[k]{l,\pathno+l-i}}}&=
      \frac{-4\PDPratio^{2+2\Pratio}-4\PDPratio^{2+\loadFactor}+
        \PDPratio^{2(\Pratio+\loadFactor)}
      +4\PDPratio^{2+2\Pratio+\loadFactor}}
      {2\PDPratio^{2+2\Pratio+\loadFactor}\loadFactor
        \left(\log{\PDPratio}\right)^3}\nonumber\\
      &+\frac{3\PDPratio^{2+2\loadFactor}-
        2\loadFactor^{\loadFactor+\Pratio+1}
        \left(\PDPratio^{\Pratio}\loadFactor+3\PDPratio
        \loadFactor+\Pratio-1\right)\log{\PDPratio}}
      {2\PDPratio^{2+2\Pratio+\loadFactor}\loadFactor
        \left(\log{\PDPratio}\right)^3}.
    \end{align}
    Making use of (\ref{eq:gammaTh}), (\ref{eq:denSI}) and (\ref{eq:prNu3rd}),
    the results (\ref{eq:thSI}) and (\ref{eq:nu3rd}) are straightforward.

  \item $\max(\Pratio, 1-\Pratio) \le \loadFactor \le 1$: substituting 
    (\ref{eq:coefHpr}) into (\ref{eq:numSI1}) and (\ref{eq:numSI2}), they 
    both yield
    \begin{align}\label{eq:prNu4th}
      \frac{1}{\stdUser[k]^4}\lim_{\pathno\rightarrow\infty}
      {\frac{1}{\pathno^2}
        \sum_{i=1}^{\pathno-1}{\coeffHsi[i]^2}\cdot
        \sum_{l=1}^{i}{\functionThetaQuad[k]{l,\pathno+l-i}}}&=
      \frac{-\PDPratio^{2+2\Pratio}-4\PDPratio^{2+\loadFactor}+
        \PDPratio^{2(\Pratio+\loadFactor)}+4\PDPratio^{2+2\Pratio+\loadFactor}}
      {2\PDPratio^{2+2\Pratio+\loadFactor}\loadFactor
        \left(\log{\PDPratio}\right)^3}\nonumber\\
      &-\frac{2\loadFactor^{\loadFactor+\Pratio+1}
        \left(\PDPratio^{\Pratio}\loadFactor+
        3\PDPratio\loadFactor+\Pratio-1\right)\log{\PDPratio}}
      {2\PDPratio^{2+2\Pratio+\loadFactor}\loadFactor
        \left(\log{\PDPratio}\right)^3}.
    \end{align}
    Making use of (\ref{eq:gammaTh}), (\ref{eq:denSI}) and (\ref{eq:prNu4th}),
    the results (\ref{eq:thSI}) and (\ref{eq:nu4th}) are straightforward.

  \item $\loadFactor=\pulseno/\pathno\ge1$: substituting 
    (\ref{eq:coefHpr}) into (\ref{eq:numSI1}) and (\ref{eq:numSI2}), they 
    both yield
    \begin{align}\label{eq:prNu5th}
      \frac{1}{\stdUser[k]^4}\lim_{\pathno\rightarrow\infty}
      {\frac{1}{\pathno^2}
         \sum_{i=1}^{\pathno-1}{\coeffHsi[i]^2}\cdot
         \sum_{l=1}^{i}{\functionThetaQuad[k]{l,\pathno+l-i}}}=
      \frac{2\PDPratio\left(\PDPratio^{2\Pratio}-1\right)
      -\left(\PDPratio^{\Pratio}+\Pratio+3\PDPratio\Pratio-1\right)
      \PDPratio^{\Pratio}\log{\PDPratio}}
           {\PDPratio^{2\Pratio+1}\loadFactor\left(\log{\PDPratio}\right)^3}.
    \end{align}
    Making use of (\ref{eq:gammaTh}), (\ref{eq:denSI}) and (\ref{eq:prNu5th}),
    the results (\ref{eq:thSI}) and (\ref{eq:nu5th}) are straightforward.

\end{itemize}

\subsection{Proof of Prop. \ref{th:loss}}\label{pr:loss}
At the Nash equilibrium, the transmit power for user $k$ when using an ARake 
receiver at the access point, $\powerStar[{k_A}]$, can be obtained from 
(\ref{eq:minimumPower}):
\begin{align}
  \powerStar[{k_A}]=\frac{1}{\channelgain[k]}\cdot
  \frac{\sigma^2\functionGamma[{\SIratio[k_A]}]}
       {1-\functionGamma[{\SIratio[k_A]}]\cdot
         \left(\SIratio[{k_A}]^{-1}+\MAIratio[{k_A}]^{-1}\right)},
\end{align}
where the subscript $A$ serves to emphasize that we are considering the case 
of an ARake, and where we have used the fact that $\hSP[k]$ is equal to the 
channel gain $\channelgain[k]=\herm{\vecpathgain[k]}\cdot\vecpathgain[k]=
\vectornorm{\vecpathgain[k]}^2$. Hence, (\ref{eq:loss}) becomes
\begin{align}\label{eq:loss2}
  \loss=
  \frac{\channelgain[k]}{\hSP[k]}\cdot
  \frac{\efficiencyFunction[{\functionGamma[{\SIratio[k_A]}]}]}
  {\efficiencyFunction[{\functionGamma[{\SIratio[k]}]}]}\cdot
  \frac{\functionGamma[{\SIratio[k]}]}
  {\functionGamma[{\SIratio[k_A]}]}\cdot
  \frac{1-\functionGamma[{\SIratio[k_A]}]\cdot
    \left(\SIratio[{k_A}]^{-1}+\MAIratio[{k_A}]^{-1}\right)}
  {1-\functionGamma[{\SIratio[k]}]\cdot
    \left(\SIratio[k]^{-1}+\MAIratio[k]^{-1}\right)}.
\end{align}

To show that \loss converges a.s. to the non-random limit of 
(\ref{eq:loss}), it is convenient to rewrite the ratio 
$\channelgain[k]/\hSP[k]$ as
\be\label{eq:PrakeRatio1}
  \frac{\channelgain[k]}{\hSP[k]}=
  \frac{\frac{1}{\pathno}\herm{\vecpathgain[k]}\cdot\vecpathgain[k]}
       {\frac{1}{\pathno}\herm{\vecrakecoeff[k]}\cdot\vecpathgain[k]}.
\ee
It is possible to prove \cite{bacci} that
\begin{align}
  \frac{1}{\pathno}\herm{\vecpathgain[k]}\cdot\vecpathgain[k] &\as 
  \functionPhi[{(\MatpathProfile[k])^2}]
  \intertext{and, analogously,}
  \frac{1}{\pathno}\herm{\vecrakecoeff[k]}\cdot\vecpathgain[k] &\as 
  \functionPhi[{\MatpathProfile[k] \MatrakeProfile[k]}].
\end{align}
Taking into account (\ref{eq:MatpathProfilePRake}), 
\begin{align}\label{eq:numPrakeRatio}
  \functionPhi[{(\MatpathProfile[k])^2}]&=
  \lim_{\pathno\to\infty}\frac{\varUser[k]}{\pathno}
  \sum_{l=1}^{\pathno}{\PDPratio^{-\frac{l-1}{\pathno-1}}}\nonumber\\
  &=\varUser[k]\cdot\frac{\PDPratio-1}{\PDPratio\log{\PDPratio}}.
\end{align}
Using (\ref{eq:denMAI1}), (\ref{eq:PrakeRatio1}) and (\ref{eq:numPrakeRatio}), 
\be\label{eq:PrakeRatio2}
  \frac{\channelgain[k]}{\hSP[k]}\as\functionMu[{\PDPratio,\Pratio}],
\ee
where $\functionMu[{\PDPratio,\Pratio}]$ is defined as in 
(\ref{eq:functionMu}).

Making use of (\ref{eq:thMAI}), (\ref{eq:thSI}), (\ref{eq:muArakeExp}),
(\ref{eq:nuArakeExp}) and (\ref{eq:PrakeRatio2}), when the hypotheses 
of Theorem \ref{th:oldPaper} hold, (\ref{eq:loss2}) converges a.s. to
(\ref{eq:loss}).

\newpage

\begin{table}
  \renewcommand{\arraystretch}{1.3}
  \caption{List of parameters used in the simulations.}
  \label{tab:parameters}
  \centering
  \begin{tabular}{c|c}
    \hline
    $\totalbits$, total number of bits per packet & $100\,\text{b}$ \\
    \hline
    $\infobits$, number of information bits per packet & $100\,\text{b}$ \\
    \hline
    $\rate[]$, bit rate & $100\,\text{kb/s}$ \\
    \hline
    $\sigma^2$, AWGN power at the receiver & 
    $5 \times 10^{-16}\,\text{W}$\\
    \hline
    $\pmax[]$, maximum power constraint & $1\,\mu\text{W}$ \\
    \hline
  \end{tabular}
\end{table}

\end{document}